\documentclass{aa}  

\usepackage{graphicx}
\usepackage{txfonts}
\usepackage{graphicx}
\usepackage{subcaption}
\usepackage[dvipsnames]{xcolor}
\usepackage{lipsum}
\usepackage{braket}
\usepackage{amsmath}
\usepackage{longtable}
\usepackage{booktabs}
\usepackage{makecell}
\usepackage{siunitx}
\usepackage{float}
\usepackage{booktabs}
\usepackage{makecell}
\usepackage{caption}
\usepackage[hidelinks,colorlinks=true,linkcolor=blue,citecolor=blue]{hyperref}
\usepackage[T1]{fontenc}
\usepackage{ae,aecompl}
\usepackage{newtxtext,newtxmath}
\usepackage{soul}

\providecommand{\sorthelp}[1]{}  
\newcommand{\dif}{\ensuremath{\mathrm{d}}}

\begin{document}

\title{The BINGO project}

\subtitle{X. Cosmological parameter constraints from HI Intensity Mapping lognormal simulations}

\author{Pablo Motta\inst{1,2,3,4}\thanks{Contact e-mail: \href{pablomotta@ustc.edu.cn}{pablomotta@ustc.edu.cn}} 
\and Filipe B. Abdalla\inst{2,3,1,5}\thanks{Contact e-mail:\href{filipe.abdalla@gmail.com}{filipe.abdalla@gmail.com}}
\and Camila P. Novaes\inst{6} 
\and Elcio Abdalla\inst{1,7,8}\thanks{Contact e-mail:\href{eabdalla@usp.br}{eabdalla@usp.br}}
\and Jiajun Zhang\inst{9,10} 
\and Gabriel A. Hoerning\inst{1,11} 
\and Alessandro Marins\inst{2,3} 
\and Eduardo J. de Mericia\inst{6} 
\and Luiza O. Ponte\inst{1} 
\and Amilcar R. Queiroz\inst{12} 
\and Thyrso Villela\inst{6,13,14} 
\and Bin Wang\inst{15,16} 
\and Carlos A. Wuensche\inst{6} 
\and Chang Feng\inst{2,3} 
\and Edmar C. Gurjão\inst{17}}

\institute{Instituto de F\'isica, Universidade de S\~ao Paulo - C.P. 66318, CEP: 05315-970, S\~ao Paulo, Brazil
\and Department of Astronomy, University of Science and Technology of China, Hefei 230026, China
\and School of Astronomy and Space Science, University of Science and Technology of China, Hefei 230026, China
\and ASTRON, the Netherlands Institute for Radio Astronomy, Oude Hoogeveensedijk 4, 7991 PD Dwingeloo, Netherlands
\and Department of Physics and Electronics, Rhodes University, PO Box 94, Grahamstown, 6140, South Africa
\and Instituto Nacional de Pesquisas Espaciais, Divisão de Astrofísica, Av. dos Astronautas, 1758, 12227-010 - São José dos Campos, SP, Brazil
\and Universidade Estadual da Paraíba, Rua Baraúnas, 351, Bairro Universitário, Campina Grande, Brazil 
\and Departamento de Física, Centro de Ciências Exatas e da Natureza, Universidade Federal da Paraíba, CEP 58059-970, João Pessoa, Brazil
\and Shanghai Astronomical Observatory (SHAO), Nandan Road 80, Shanghai, 200030, China
\and State Key Laboratory of Radio Astronomy and Technology, 20A Datun Road, Chaoyang District, Beijing, 100101, China
\and Jodrell Bank Centre for Astrophysics, Department of Physics \& Astronomy, The University of Manchester, Oxford Road, Manchester, M13 9PL, U.K.
\and Unidade Acadêmica de Física, Univ. Federal de Campina Grande, R. Aprígio Veloso, 58429-900 - Campina Grande, Brazil.
\and Centro de Gestão e Estudos Estratégicos, SCS, Qd. 9, Lote C, Torre C S/N, Salas 401 a 405, 70308-200 - Brasília, DF, Brazil.
\and Universidade de Brasília, Instituto de Física, Campus Universitário Darcy Ribeiro, 70910-900 - Brasília, DF, Brazil.
\and Center for Gravitation and Cosmology, Yangzhou University, Yangzhou 224009, China
\and School of Aeronautics and Astronautics, Shanghai Jiao Tong University, Shanghai 200240, China
\and Unidade Acadêmica de Engenharia Elétrica, Univ. Deferal de Campina Grande, R. Aprígio Velos, 58429-900 - Campina Grande, Brasil
}

   \date{Received XXX XX, XXXX; accepted XXX XX, XXXX}


  \abstract
   {Building on the transformative success of optical redshift surveys, the emerging technique of neutral hydrogen (HI) intensity mapping (IM) offers a novel probe of large-scale structure (LSS) growth and the late-time accelerated expansion of the universe.}
   {We present cosmological forecasts for the Baryon Acoustic Oscillations from Integrated Neutral Gas Observations (BINGO), a pioneering HI IM experiment, quantifying its potential to constrain the \textit{Planck}-calibrated $\Lambda$CDM cosmology and extensions to the $w_0w_a$CDM dark energy model.}
   {For BINGO's Phase~1 configuration, we simulate the HI IM signal using a lognormal model and incorporate three dominant systematics: foreground residuals, thermal noise, and beam resolution effects. Using Bayesian inference, we derive joint constraints on six cosmological parameters ($\Omega_b h^2$, $\Omega_c h^2$, $100\theta_s$, $n_s$, $\ln 10^{10} A_s$, and $\tau_r$) alongside 60 HI parameters ($b_{\rm HI}^i$, $\Omega_{\rm HI}^i b_{\rm HI}^i$) across 30 frequency channels.}
   {Our results demonstrate that combining BINGO with the Planck 2018 CMB dataset tightens the confidence regions of cosmological parameters to $\sim$40\% the size of those from Planck alone, significantly improving the precision of parameter estimation. Furthermore, BINGO constrains the redshift evolution of HI density and delivers competitive measurements of the dark energy equation of state parameters ($w_0$, $w_a$).}
   {These results demonstrate BINGO’s potential to extract significant cosmological information from the HI distribution and provide constraints competitive with current and future cosmological surveys.}


   \maketitle

%

\section{Introduction}

One of the major challenges in modern cosmology is understanding the late-time accelerated expansion of the Universe. Such an expansion is driven by Dark Energy (DE), an exotic form of matter with negative pressure. In spite of two decades of research, a cosmological constant $\Lambda$ remains the prevailing DE candidate. DE, together with Cold Dark Matter (DM) - such an unknown physical component is responsible for galaxy clusters and large-scale structures - compose the $\Lambda$CDM (cosmological) model. The $\Lambda$CDM model has successfully been supported by numerous astronomical observations. Examples are galaxy photometric surveys such as KiDS \citep{hildebrandt2017kids}, DES \citep{abbott2022dark}, Cosmic Microwave Background (CMB) fluctuations \citep{planck2016-l01}, type Ia Supernovae \citep{scolnic2018complete} and galaxy cluster counts \citep{de2016cosmological}, among others.

However, the $\Lambda$CDM model suffers two serious theoretical
problems. We cite the cosmological constant problem \citep{weinberg1989cosmological} and the coincidence problem \citep{Wang:2016lxa,Wang:2024vmw,chimento2003interacting}. Also, there are some observational discrepancies such as the $H_0$ tension \citep{riess20113, riess20162} and the $\sigma_8$ tension \citep{ade2016astron, hamann2013new, battye2014evidence, petri2015emulating}. Other probes provide additional information about the Universe's evolution and are essential to indicate whether alternatives to the $\Lambda$CDM model are more suitable to explain the current observations \citep{Costa:2018aoy,abdalla2020dark, Hoerning:2023hks}. 

The standard approach to probe the Large-Scale Structure (LSS) is through large galaxy-redshift surveys, such as the Two-degree-Field Galaxy Redshift Survey \citep[2dFGRS;][]{colless20012df}, the WiggleZ Dark Energy Survey \citep{blake2011wigglez, kazin2014wigglez}, the Six-degree-Field Galaxy Survey \citep[6dFGS;][]{jones20096df, beutler20116df} and the Baryon Oscillation Spectroscopic Survey \citep[BOSS;][]{ross2012sdss}, which is part of the third stage of the Sloan Digital Sky Survey \citep{york2000sloan, anderson2012clustering, alam2017clustering}. The Dark Energy Survey (DES) recently reported their cosmological constraints with the 3-year data \citep{troxel2018dark, camacho2019dark}. Future optical surveys that aim to use larger and more sensitive telescopes at a variety of high redshifts, such as the Dark Energy Spectroscopic Instrument \citep[DESI][]{levi2013desi, aghamousa2016desi}, Large Synoptic Survey Telescope \citep[LSST][]{ivezic2019lsst, ivezic2009lsst, mandelbaum2018lsst}, Euclid \citep{laureijs2011euclid}, WFIRST \citep{green2012wide}, and CSST \citep{gong2019cosmology, miao2023cosmological} have been proposed, and some of the constructions are underway. DESI delivered some interesting results which may point to new physics \cite{karim2025desi} concerning a departure from standard cosmology (see also  \cite{Wang:2005jx}) To date, galaxy-redshift surveys have made significant progress in studying the LSS of the Universe. To perform precision cosmology, one must detect sufficiently large samples of neutral hydrogen (HI) emitting galaxies. However, this is a huge task since at higher redshifts, and the galaxies look essentially very faint \citep{bull2015late, kovetz2017line, pritchard201221}.

Another method for probing LSS is the Intensity Mapping (IM) technique, which is less expensive and more efficient. Instead of measuring individual galaxies, IM measures the total flux from many galaxies within a beam resolution, which will underline matter. In its most common variant, 21-cm Intensity Mapping, the neutral hydrogen emission line is used as a tracer of matter. There are two main reasons to use IM instead of galaxy surveys: i) a larger volume of the Universe can be surveyed in less time, ii) the redshift is measured directly from the redshifted 21-cm line \citep{olivari2018intensity}. This is the case of the Baryon Acoustic Oscillations from the Integrated Neutral Gas Observations (BINGO) experiment \citep{abdalla2022bingo}, which is projected, among other things, to measure Baryon Acoustic Oscillations (BAO) in the radio band.

The HI IM also faces some challenges from astrophysical contamination and systematic effects. At the frequency window $\sim$ 1 GHz, HI signals are observed together with other astrophysical emissions due to the type of observation generically called foreground emissions, such as the Galactic synchrotron radiation and extra-galactic point sources \citep{battye2013h}. The signal level of HI IM (T $\sim$ 1 mK) is about 4 orders of magnitude lower than the foreground emissions (T $\sim$ 10 K). Thus, preparing an efficient foreground removal technique to estimate the signal from contamination is crucial to the HI IM probes \citep[e.g.,][]{wolz2014effect, alonso2015blind, olivari2016extracting}. On the other hand, systematic effects, primarily related to the instrument, may behave similarly to the signal or even cover it at small scales (e.g., the 1/f noise). Besides, they will also make the foreground removal procedure harder. A careful treatment of systematic effects is hence necessary for HI IM experiments. The Generalized Needlet Internal Linear Combination ({\tt GNILC}) is a component separation developed by \cite{remazeilles2011foreground} and successfully used in the Planck data analysis effort \citep[see, e.g.,][]{aghanim2016planck}. It is one of the algorithms used to estimate the foreground contribution to the measurements made by BINGO collaboration \citep{olivari2016extracting, liccardo2022bingo, fornazier2022bingo, marins2022foreground, de2023testing}.

Several experiments have already demonstrated the feasibility of 21\,cm intensity mapping as a tool for cosmology. Early detections were achieved by the Green Bank Telescope (GBT) through cross-correlation with optical galaxy surveys, yielding measurements of large-scale structure at redshifts around $z \sim 0.8$~\citep{chang2010intensity, masui2013measurement}. The Parkes telescope also detected the 21\,cm signal in cross-correlation with 2dF galaxies at lower redshifts~\citep{anderson2018low}, further validating the technique. More recently, instruments like the CHIME Pathfinder have shown promising results in the development of large-scale intensity mapping surveys~\citep{bandura2014canadian}, while MeerKAT is being employed in single-dish mode to test auto- and cross-correlation methods for cosmological signal extraction~\citep{cunnington2019impact, meerklass2025meerklass}. Building upon these successful detections, a number of ongoing and upcoming telescopes are designed to probe the 21\,cm signal over a wide redshift range, with the goal of detecting baryon acoustic oscillations (BAOs) and constraining cosmological parameters. These include the Canadian Hydrogen Intensity Mapping Experiment~\citep[CHIME;][]{bandura2014canadian}, the Five-hundred-meter Aperture Spherical Radio Telescope~\citep[FAST;][]{nan2011five, bigot2015hi}, the Square Kilometre Array~\citep[SKA;][]{santos2015cosmology}, and Tianlai~\citep{chen2012tianlai}. The BINGO project is part of this new generation of experiments and aims to make precise measurements of the 21\,cm signal at intermediate redshifts ($z \sim 0.13$–$0.45$), enabling competitive constraints on late-time cosmological parameters, including those related to dark energy.

Several aspects of the BINGO project, including instrument
description, component separation techniques, simulations, forecasts for cosmological models, and BAO signal recoverability and search for Fast Radio Bursts are found in the series of papers \citep{abdalla2022bingo, wuensche2022bingo, abdalla2022bingo-iii, liccardo2022bingo, fornazier2022bingo, costa2022bingo, novaes2022bingo, santos2023bingo} as well as in \cite{xiao2021forecasts, marins2022foreground, de2023testing, novaes2024cosmological}. 

In this paper, we propose cosmological parameter constraints from HI Intensity Mapping simulations for BINGO phase 1 operation. The simulations include the cosmological 21-cm signal, assumed as a multivariate lognormal distribution, the main foreground sources contributing
to the BINGO frequency range, the instrumental noise, as well as the effect introduced by a fixed instrumental beam resolution. We apply the angular power spectrum (APS) 
formalism to the BINGO simulations and follow a standard
Bayesian analysis framework in order to obtain constraints for  BINGO and BINGO + Planck 2018.

This work is organized as follows. In section \ref{sec.theo_cl} we derive the angular power spectrum ($C_\ell$) of the brightness temperature of the IM 21-cm signal. Section \ref{sec.BINGO_sim} describes the BINGO simulations. Sections \ref{sec.pseudo_aps_estimate} and \ref{sec.covariance_matrix} describe our pseudo-$C_\ell$ estimate and covariance matrix computation. In section \ref{sec.cosmological_analysis} we explain the Bayesian modeling for cosmological parameter estimation, and finally, section \ref{sec.results} presents the results of cosmological analysis. We assessed the impact of various error sources on the accuracy of our cosmological parameter estimates and also compared constraints derived from different realizations of the
BINGO simulation. 

\section{The $C_\ell$ of the HI brightness temperature}
\label{sec.theo_cl}

In intensity mapping experiments, the brightness temperature of the HI (neutral hydrogen) line is used to study the large-scale structure of the Universe. Such emission results from a transition between two energy levels of the hydrogen atom, corresponding to the hyperfine structure. The emission in the rest frame has a wavelength of 21-cm and the corresponding frequency of $\nu_{10} = 1420 \text{MHz}$. Following \cite{furlanetto2006cosmology,pritchard201221}, the observed (average) 21-cm brightness temperature at low-redshifts, $z < 2$, is 
 \begin{align}
     \Bar{T}_{\rm HI}(z) = \Big( \frac{9 \hbar c^3A_{10} }{128 \pi Gk_B \nu_{10} m_{\rm HI} } \Big) \frac{\Omega_{\rm HI}(z) }{(1+z)^2} \frac{H_0^2}{ || \dif v / \dif \chi || } \,, 
     \label{eq.brightness_temp_1}     
 \end{align}
where $H_0$ is the Hubble constant and $|| \dif v / \dif \chi ||$ is the gradient of the specific velocity field along the line of sight and  $\Omega_{\rm HI}(z)$ is the density parameter of HI.  $G$, $c$, $k_B$, and $\hbar$ are the gravitational constant, speed of light, Boltzmann constant, and reduced Planck constant, respectively. $A_{10}$ represents the spontaneous emission coefficient of the 21-cm transition, and $m_{\rm HI}$ is the HI atom mass. Following \cite{battye2013h, hall2013testing} this expression gets simplified by referring to the cosmic background evolution, where we can express the gradient of the specific velocity only in terms of Hubble flow
\begin{align}
    \Big|\Big| \frac{ \dif v }{ \dif \chi } \Big|\Big| = \frac{H(z) }{ (1+z)^4 },\quad .
\end{align}
Replacing the values of the fundamental constants, we find
 \begin{align}
     \Bar{T}_{\rm HI}(z) = 188 h \, \Omega_{\rm HI}(z) \frac{(1+z)^2}{E(z)} \, \text{mK} \,, 
     \label{eq.brightness_temp_2}
 \end{align}
where $H_0 = 100\, h \,\, {\rm km \,\, s}^{-1} {\rm Mpc}^{-1}$ and $E(z) = H(z)/H_0$.

Like the dark matter distribution, the brightness temperature is not homogeneously distributed, and by considering its spatial anisotropy, $\Delta T_{\rm HI}(z, \hat{n})$, we depict the temperature filed as $T_{\rm HI} ( z, \hat{n} ) = \bar{T}_{\rm HI} (z) + \Delta T_{\rm HI} (z, \hat{n})$, where $\hat{n}$ is the unit vector along the line of sight (LoS). The temperature fluctuation $\Delta T_{\rm HI}$ results from various mechanisms during the cosmic evolution, including the foremost contribution from the HI overdensity $\delta_{\rm HI}$, and approximately we have $\Delta T_{\rm HI} = \bar{T}_{\rm HI} \delta_{\rm HI} $.

Since 21-cm IM surveys the Universe within tomographic frequency channels (or equivalently, frequency bins), the observable in practice is a projection of the 3D temperature field along LoS, such that the integrated temperature fluctuation reads
\begin{align}
    \delta T_{\rm HI} (\hat{n}) = \int \dif z \, \phi(z) \Bar{T}_{\rm HI}(z) \delta_{\rm HI} ( z, \hat{n} ),
\end{align}
where $\phi(z)$ is the window function determining the bin width. Assuming a top-hat filter, we have $\phi(z) = 1/( z_{max} - z_{min} )$ for $z_{min} < z < z_{max} $ and $\phi(z) =0$ outside the frequency channel \citep{battye2013h}. 

Following \cite{novaes2022bingo}, the resulting angular power spectrum $C_\ell^{ij} = \langle a_{\ell m}^i a_{\ell m}^{j*}  $ for the projected HI brightness temperature is
\begin{align}
    C_\ell^{ij} = \frac{2}{\pi} \int \dif k W^i_{HI,\ell} (k)W^j_{HI,\ell}(k) k^2 P(k) \,.
    \label{eq.cl_uclcl}
\end{align}
where the indices $i,j$ denote tomographic frequency channels such that, for $i = j$ and $i \neq j$ we obtain auto- and cross-APS, respectively. 
The $W^i_{HI,\ell}$ term is a window function collecting all the redshift dependencies
\begin{align}
    W^i_{HI,\ell}(k) = \int \dif z \, b_{\rm HI}(z) \ \phi (z) \bar{T}_{\rm HI}(z)D(z) j_\ell (k\chi(z)) \,,
    \label{eq.window_func_hi}
\end{align}
where $D(z)$ is the growth function, $b_{\rm HI} (z)$ is the HI bias as a function of the redshift, and $j_\ell$ is the spherical Bessel function. 
Equation (\ref{eq.window_func_hi}) depends on two parameters of the HI astrophysics, $b_{\rm HI}(z)$ and $\Omega_{\rm HI}(z)$ (through Eq. (\ref{eq.brightness_temp_2})). For simplicity, we assume that the frequency channels are sufficiently thin, such that both parameters can be assumed to have constant values inside each bin. We also assume them to be scale-independent.

In our work, we use the {\tt UCLCl} code \citep{mcleod2017joint,loureiro2019cosmological} a library for calculating the angular power spectrum of the cosmological fields. Such code obtains the primordial power spectra and transfer functions from the \texttt{CLASS} Boltzmann code \citep{blas2011cosmic} and applies Eq. \ref{eq.cl_uclcl} to obtain the APS. 
In the \texttt{UCLCl}, the $C_\ell$ computation may be extended some way into the non-linear regime by introducing the scale-dependent non-linear overdensity $\delta_{NL} (k,\chi)$ from the \texttt{CLASS} code \citep{blas2011cosmic, di2013classgal}, using the modified \textsc{halofit} of \cite{takahashi2012revising}. See \cite{loureiro2019cosmological} for details. 

\subsection{Redshift space distortion}

Galaxies have peculiar velocities relative to the overall expansion of the Universe. This creates the illusion that the local peculiar motion of galaxies moving towards us makes them appear closer (i.e., appear to be at lower redshifts), while galaxies with peculiar motion moving away from us appear to be further away (i.e., they appear to be at higher redshifts). This effect is called by \emph{Redshift space distortion} \citep[RSD;][]{kaiser1987clustering}, and it is also taken into account by the {\tt UCLCl} code. 

In this work, we follow the same formalism shown in \cite{fisher1994spherical, padmanabhan2007clustering}. This formalism was presented for galaxy probes, and we adapted it for our Intensity Mapping window function. Assuming that the peculiar velocities are small compared with the thickness of the redshift slice, we write $\bar{T}_{\rm HI}$ as a function of the redshift distance $s = \chi + \mathbf{v} \cdot \hat{n}$ and take the first order Taylor expansion
\begin{align}
    \bar{T}_{\rm HI} (s) = \bar{T}_{\rm HI} (\chi) + \frac{\dif \bar{T}_{\rm HI}}{\dif \chi} \cdot (\mathbf{v}(\chi\hat{n})\cdot\hat{n} ) \,.
\end{align}
The first order term leads to a correction on the window function used in Eq. \ref{eq.cl_uclcl}, which becomes $W^i_{Tot,\ell} = W^i_{HI,\ell} + W^i_{RSD,\ell}$, with the RSD contribution given by 
\begin{align}
    W^i_{RSD,\ell}(k) &= \frac{\beta^i}{k} \int \dif\chi \phi^i (\chi) \frac{\dif \bar{T}_{\rm HI}}{\dif \chi} j'_\ell( k \chi ) \\
    &= \beta^i  \int \dif \chi \, \phi^i (\chi) T_{\rm HI}(\chi) \Big[   \frac{ (2\ell^2 + 2\ell + 1) }{(2\ell+3)(2\ell-1)} j_\ell ( k\chi ) -
    \nonumber\\  &  
    -\frac{\ell(\ell-1)}{(2\ell-1)(2\ell+1)}j_{\ell -2} ( k\chi ) -
    \nonumber\\  & -
    \frac{(\ell+1)(\ell+2)}{(2\ell+1)(2\ell+3)}j_{\ell +2} ( k\chi ) \Big] \,,
\end{align}
where $\beta^i$ is the redshift distortion parameter, $ \beta^i =
(\dif \ln D(z)/ \dif \ln a)/ b^i_{\rm HI}$. The RSD window function does not consider the Fingers of God effect, which affects small scales due to the virial motion of galaxies inside clusters \citep{kang2002analytical}.

\subsection{Partial sky: mixing matrix convolution}

The modeling given by Eq. (\ref{eq.cl_uclcl}) assumes the (brightness) temperature as an isotropic random field defined in the whole sphere. However, for comparison with the data, we need to take into account that we only survey a portion of the sky.  The angular mask function $W(\mathbf{\hat{n}})$ can be expanded into spherical harmonics coefficients $w_{\ell m} = \int \dif \mathbf{\hat{n}} W(\mathbf{\hat{n}}) Y_{\ell m}^* (\mathbf{\hat{n}}) $ and with a power spectrum
\begin{align}
    \mathcal{W}_\ell = \frac{1}{2\ell+1} \sum_m |w_{\ell m}|^2.
\end{align}
Following \cite{hivon2002master} and \cite{blake2007cosmological}, the effect of the angular mask function on the power spectrum is given by a convolution with a coupling kernel
\begin{align}
    S_{\ell} = \sum_{\ell'} R_{{\ell}{\ell'}} C_{\ell'} \, ,
    \label{eq.PseudoAPS}
\end{align}
where $R_{{\ell}{\ell'}}$, named mixing matrix, is given by
\begin{align}
    R_{\ell \ell'} =  \frac{(2\ell'+1)}{4\pi f_{sky} } \sum_{\ell''}  \mathcal{W}_{\ell''} 
    \begin{pmatrix}
    \ell & \ell' & \ell'' \\
    0 & 0 & 0 
    \end{pmatrix}^2 \,.
    \label{eq.mixing_matrix_renorm}
\end{align}
The mixing matrix depends only on the mask function given by its angular power spectrum $\mathcal{W}_\ell$. The $2 \times 3$ matrix above is the Wigner $3j$ function; these coefficients were calculated using the UCLWig 3j library\footnote{\url{https://github.com/LorneWhiteway/UCLWig3j}}, which optimises the calculation of Wigner $3j$ symbols using the recurrence
relation by \cite{schulten1976recursive}.

\section{BINGO simulations}
\label{sec.BINGO_sim}

\begin{figure}
    \centering
    \includegraphics[scale=0.55]{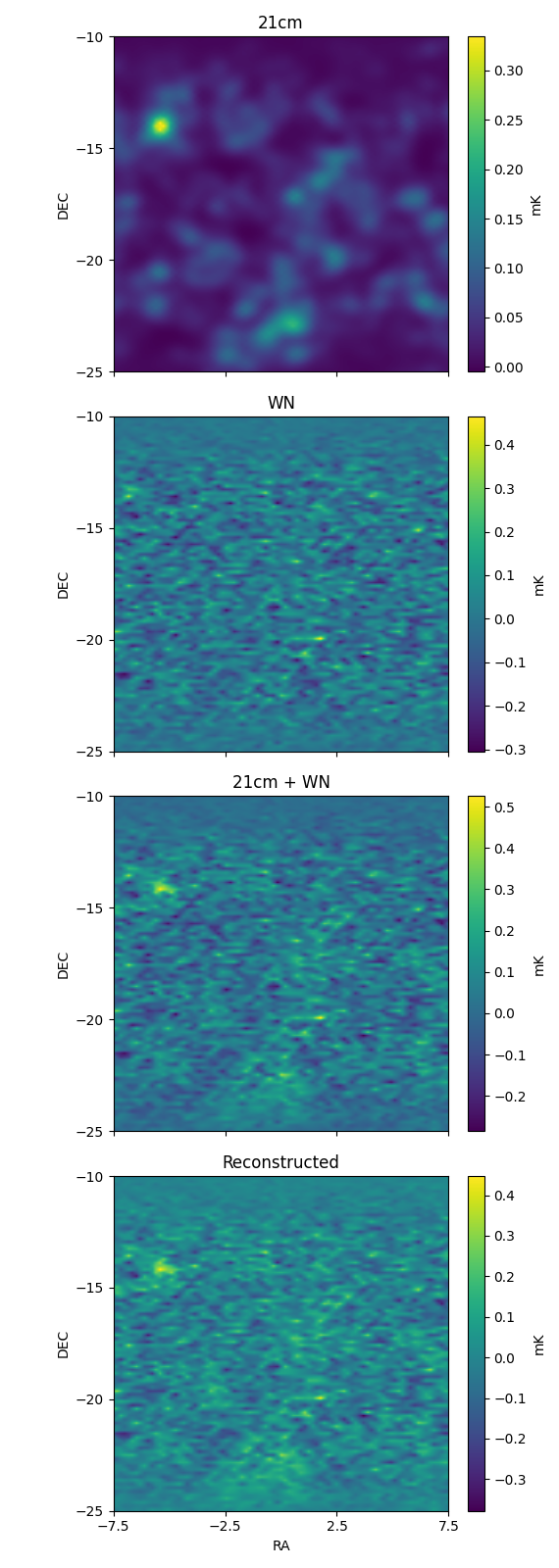}
    \caption{From top to bottom, each panel shows the pure 21-cm cosmological signal generated with the {\tt FLASK} code (subsection \protect\ref{sec.cosmological_signal}) convolved with a Gaussian beam of $\theta_{\rm FWHM} = 40$ arcmin, a realization of the white noise contribution (subsection \protect\ref{sec.instrumental}), the cosmological signal and white noise summed up and the reconstructed map using FastICA.}
    \label{fig.bingo_sim}
\end{figure}

\begin{table}
\scriptsize
\caption{Summary of the BINGO telescope and survey parameters (Phase 1).}
\label{tab:bingo_forecast_params}
\centering
\begin{tabular}{lrl}
\hline\hline
Parameter & Value & Unit \\
\hline
\multicolumn{3}{c}{\textbf{Receiver and Frequency Band}} \\
System temperature, $T_{\rm sys}$                    & 70                    & K \\
Frequency range ($f_{\rm min}$ – $f_{\rm max}$)      & 980 – 1260           & MHz \\
Bandwidth, $B$                                       & 280                  & MHz \\
Number of redshift bins\tablefootmark{a}            & 30                   & \\
Sampling time, $\tau_s$                              & 10                   & Hz \\
Estimated sensitivity (1 yr, 1 pixel)\tablefootmark{b}   & 206                  & $\mu$K \\
Instrument noise (1 s, 1 $z$-bin)\tablefootmark{c}       & 23                   & mK \\
Circular polarizations                               & 2                    & \\
\hline
\multicolumn{3}{c}{\textbf{Telescope Geometry}} \\
Instantaneous field of view\tablefootmark{d}         & $14.75 \times 6.0$    & deg$^2$ (Dec, RA) \\
Number of horns (Phase 1)                            & 28                   & \\
Telescope beam solid angle ($\Omega_{\rm beam}$)\tablefootmark{e}   & 0.35  & deg$^2$ \\
Optics FWHM (at 1.1 GHz)                              & $\sim$0.67           & deg \\
\hline
\multicolumn{3}{c}{\textbf{Collecting Area}} \\
Geometrical area, $A_{\rm tot}$                      & 1602                 & m$^2$ \\
Effective area, $A_{\rm eff}$\tablefootmark{f}       & $\sim$1120           & m$^2$ \\
Mean $A_{\rm eff}$ from GRASP\tablefootmark{f}       & 621                  & m$^2$ \\
\hline
\multicolumn{3}{c}{\textbf{Survey Strategy}} \\
Maximum instantaneous declination coverage\tablefootmark{g} & 14.75           & deg \\
Maximum declination range\tablefootmark{g}           & 15.31                & deg \\
Full survey area (5 years)                            & $\sim$5324           & deg$^2$ \\
Mission duration                                      & 5                    & years \\
Duty cycle (estimated)                                & 60–90                & \% \\
\hline
\end{tabular}
\tablefoot{\scriptsize
\tablefoottext{a}{Assumed for science forecasts; actual operational binning is derived from the digital backend output.} \\
\tablefoottext{b}{Assuming 1 pixel, full band, 1 year, and 60\% duty cycle.} \\
\tablefoottext{c}{Assuming 1 redshift band, 1 polarization, 28 horns, and 60\% duty cycle.} \\
\tablefoottext{d}{Field of view measured horn-to-horn.} \\
\tablefoottext{e}{Beam solid angle computed at 1120 MHz.} \\
\tablefoottext{f}{Mean $A_{\rm eff}$ estimated from GRASP simulations; actual values vary between 532–653 m$^2$.} \\
\tablefoottext{g}{Declination range defined by horn height and fixed mirror geometry.} \\
}
\end{table}

The simulations used in our analysis are designed to reproduce the observational configuration of the BINGO Phase~1 experiment, with parameters summarized in Table~\ref{tab:bingo_forecast_params}. Each simulated map includes the cosmological 21-cm signal, thermal instrumental noise, and the beam smoothing effect, modeled as a symmetric Gaussian with $\theta_{\rm FWHM} = 40$~arcmin, consistent with BINGO’s optical design. The full BINGO frequency band, spanning 980–1260~MHz, is divided into 30 tomographic bins of $\Delta \nu \approx 9.33$~MHz, matching the resolution of the digital backend. Simulations are produced using the HEALPix pixelization scheme \citep{gorski2005healpix} with $N_{\rm side} = 256$, which provides an angular resolution well-suited to the telescope’s beam size.

We generate 3000 independent realizations of the cosmological 21-cm signal. To each realization, we add one instance of thermal (white) noise and simulated astrophysical foregrounds to create a set of contaminated total sky maps. We then apply the FastICA component separation algorithm to each of these 3000 maps to perform blind foreground removal. This process yields a corresponding set of 3000 reconstructed maps, which contain the recovered 21-cm signal and instrumental noise. One randomly selected realization from this set is used as the mock data vector for cosmological inference, while the full ensemble is used to compute the covariance matrix. An example of each simulation component and their combination is shown in Figure~\ref{fig.bingo_sim}.

\subsection{\bf Cosmological signal.}
\label{sec.cosmological_signal}

We used the publicly available {\tt FLASK}\footnote{\url{http://www.astro.iag.usp.br/~flask/}} code \citep{2016_xavier} to produce two-dimensional tomographic realizations (spherical shells around the observer, positioned at the center) of random astrophysical fields following a multivariate lognormal distribution, reproducing the desired cross-correlations between them. The lognormal distribution is more suitable to describe matter tracers than the Gaussian distribution, avoiding non-physical ($\delta < -1$) densities. 
 {\tt FLASK} takes as input the auto- and cross-$C^{ij}_\ell$ ($i=j$ and $i \neq j$, respectively) previously calculated for each of the $i,j$ redshift slices. 
The 21\,cm $C^{ij}_\ell$ are computed with the {\tt UCLCl} code using the Planck 2018 $\Lambda$CDM fiducial cosmology \citep[$\Omega_b = 0.0493 $, $\Omega_{c} = 0.2645 $,  $h =0.6736 $, $n_s = 0.9649 $, $\ln 10^{10} A_s = 3.044 $, $\tau_r = 0.0544 $;][]{planck2016-l06} and HI parameters given by $\Omega_{\rm HI}= 3.642 \times 10^{-4}$ \citep{padmanabhan2015theoretical} and $b_{\rm HI}=1$. From them, we generated 3000 lognormal realizations, each of them corresponding to 30 {\tt HEALPix} full sky maps of 21-cm brightness temperature fluctuations, one for each BINGO frequency (or redshift) bin. For simplicity, and following some of the previous works of the BINGO collaboration \citep{zhang2022bingo, fornazier2022bingo}, we fix the number of channels at 30 ($\delta \nu = 9.33$ MHz). The impact of different tomographic binning for the cosmological analyses will be evaluated in the future \citep[see][for analyses evaluating the impact of the number of frequency channels in the foreground cleaning process]{de2023testing}. 

\subsection{\bf Foreground signals.}
\label{sec.foreground_signals}
Using the Planck Sky Model software \citep[PSM;][]{2013/delabrouille}, we simulate the foreground signal contributing in each frequency bin of the BINGO band. Following \cite{novaes2022bingo} and \cite{de2023testing}, we consider the contribution of seven foreground components. 
From our Galaxy, we account for the synchrotron and free-free emissions, which are the most important contaminants in the BINGO band, as well as the thermal dust and anomalous microwave emissions. 
Among the extragalactic components, we include the thermal and kinetic Sunyaev-Zel'dovich effects and the unresolved radio point sources. 
For a detailed description of the specific configuration adopted in the PSM code to simulate each foreground component, we refer the reader to \cite{de2023testing}.

\subsection{\bf Instrumental effects and sky coverage.}
\label{sec.instrumental}

The effect of the BINGO beam is incorporated into all frequency maps by convolving them with a symmetric Gaussian profile. We adopt a fixed full-width at half maximum of $\theta_{\rm FWHM} = 40$ arcmin, which corresponds to the beam size at 1100,MHz. While the true beam width varies slightly with frequency, this variation is slow across the 980–1260,MHz BINGO band. Therefore, using a constant beam size across all channels is a good approximation for the purposes of our analysis. Moreover, the Gaussian profile has been shown to accurately reproduce the BINGO beam shape at the relevant frequencies, as demonstrated in \cite{abdalla2022bingo-iii}. We then add to the simulations the contribution of the thermal (white) noise, taking into account the BINGO specifications, as detailed in \cite{wuensche2022bingo}; \cite{abdalla2022bingo-iii}. We note that this noise model does not include $1/f$ (correlated) noise, which is a common simplification for preliminary analyses. While this omission may underestimate noise correlations at large angular scales, we expect the impact to be subdominant for BINGO's primary science goals given the scanning strategy and data processing pipelines. We assume BINGO phase 1 operation, with temperature system $T_{sys} = 70$\,K, operating with 28 horns and optical arrangement as designed in \cite{abdalla2022bingo-iii}, and considering 5 years of observation. To allow a more homogeneous coverage of the BINGO region, the horns will have their positions shifted by a fraction of beam width in elevation each year. Taking these specifications into account, the thermal noise level, or the root mean square (RMS) value, by pixel, at the resolution of $N_{side} = 256$, is estimated as described in \cite{fornazier2022bingo}. From this RMS map, we are able to generate an arbitrary number of noise realizations by multiplying it by Gaussian distributions of zero mean and unitary variance. 
In order to reproduce BINGO sky coverage, we also apply a cut sky mask to the simulations. Besides selecting BINGO area, this mask also cuts out the 20\% more contaminated Galactic region. It is also apodized with a cosine square (C2) transition, using the {\tt NaMaster} code\footnote{\url{https://namaster.readthedocs.io}} \citep{2019_alonso-namaster}, to avoid boundary artifacts when calculating the APS. Details on how this mask is constructed can be found in \cite{de2023testing}.

\subsection{\bf Foreground cleaning.} 
\label{sec.foreground_cleaning} 

Our methodology builds upon the BINGO collaboration's framework, where previous studies have extensively developed and compared blind component separation methods such as GNILC \citep{fornazier2022bingo, liccardo2022bingo, de2023testing} and FastICA \citep{marins2022foreground}. Our current implementation uses the FastICA algorithm \citep{hyvarinen1999fast}, which is based on the principle of statistical independence between components. This choice is motivated by its performance in recovering the 21-cm signal with an accuracy equivalent to other blind methods, combined with a favorable computational efficiency for processing large ensembles of simulations in the current stage of our pipeline analysis \citep{marins2022foreground}. 

We performed blind foreground removal using the {\tt Fast-ICA} algorithm implemented in the scikit-learn package\footnote{\url{https://scikit-learn.org/stable/modules/generated/sklearn.decomposition.
FastICA.html}}. The method was applied directly to our multi-frequency simulated maps. Operating without prior templates, FastICA identifies and isolates the dominant, statistically independent foreground components. After subtracting these components from the data, the residual signal is projected back into frequency space to yield a set of foreground-cleaned maps. For our analysis of 3000 sky simulations, we configured the algorithm to estimate 4 independent components. Following the setup successfully applied in similar 21-cm intensity mapping contexts\citep{marins2025investigating}, we use the 'logcosh' nonlinearity with 20 interactions, and a convergence tolerance of 0.01.

\section{APS measurements from maps}
\label{sec.pseudo_aps_estimate}

\begin{figure}
    \centering
    \includegraphics[scale=0.55]{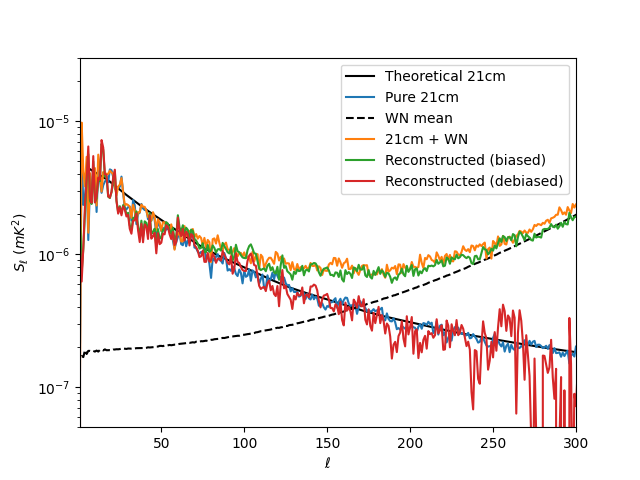}
    \caption{Comparison of angular power spectra $S_\ell$ for different simulated components. The black solid line shows the theoretical 21-cm spectrum from \texttt{UCLCl}. Colored lines represent measured spectra from simulations: pure 21-cm signal (blue), the sum of 21-cm and white noise (orange), the reconstructed spectrum after foreground removal (green, biased), and the final debiased spectrum (red) after applying the correction from Eq.~\ref{eq.debiasing}. }
    \label{fig.cl_measurements}
\end{figure}

\begin{figure*}
    \centering
    \includegraphics[scale=0.5]{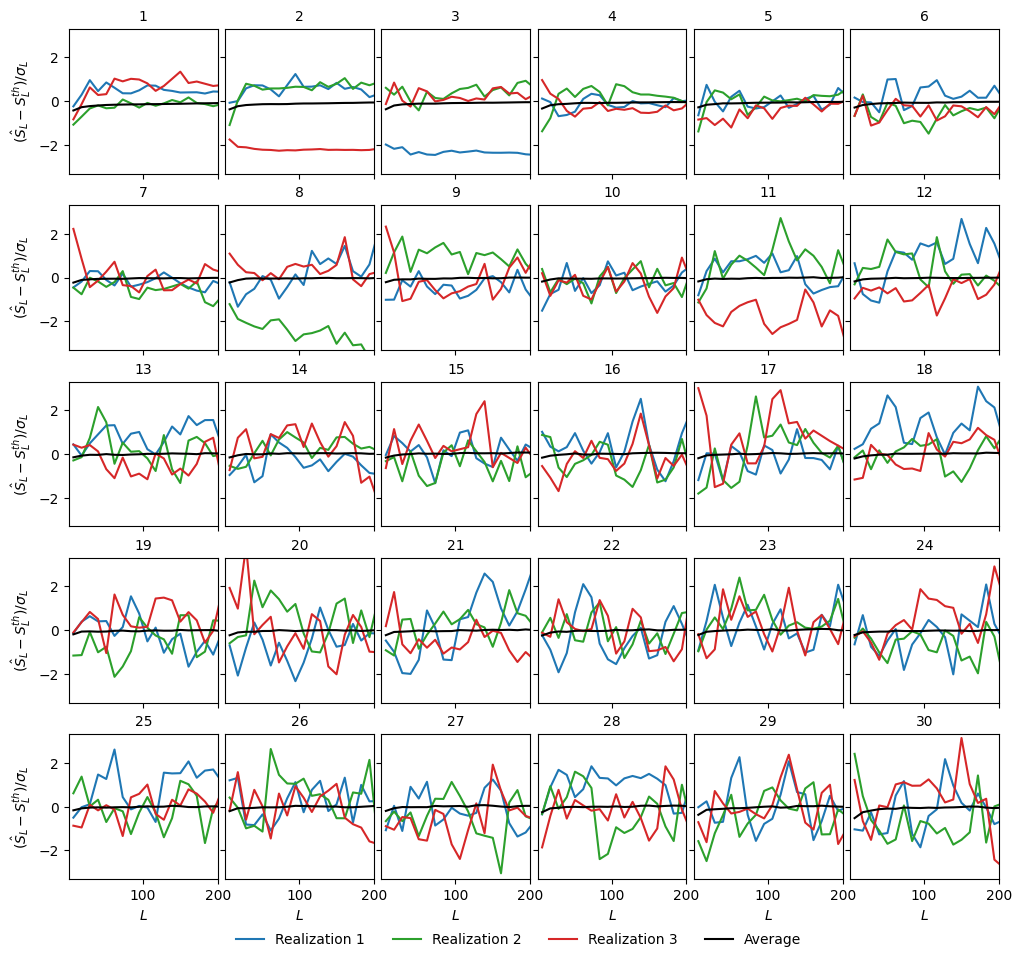}
    \caption{The blue, green, and red lines represent the relative error $(\hat{S}_\ell - S^{\mathrm{th}}_\ell)/\sigma_\ell$ computed for three different realizations of the BINGO simulations including White Noise and foreground removal. $\sigma_\ell$ is the standard deviation derived from the covariance matrix, as described in Sec.~\ref{sec.covariance_matrix}. The black line shows the average relative error over 3000 simulations. This plot illustrates that neighboring multipoles are highly correlated. Each panel corresponds to a different frequency channel, as defined in Table~\ref{tab.multipole_selection}, with the channel number indicated in the panel title.}
    \label{fig.relative_error}
\end{figure*}

Given our simulated temperature fluctuation maps, let $a_{\ell m}$ be the coefficients of spherical harmonic expansion. The pseudo-$C_\ell$ estimate that takes in account beam resolution, pixelation scheme, and sky coverage corrections is defined by
\begin{align}
    \hat{S}^{ij}_\ell = \frac{1}{f_{sky} b_\ell^2 w_\ell^2  } \sum_{m=-\ell}^\ell a_{\ell m} a_{\ell m}^*
    \label{eq.pseudo_aps_estimate}
\end{align}
Here, $b_\ell$ is the axisymmetric beam window function, e.g., \cite{challinor2011linear}, assumed in this work as a Gaussian beam and $w_\ell$ is the pixel window function \citep{gorski2005healpix, leistedt2013estimating}. The beam window function corrects the effect of the beam resolution on the observed angular power spectrum. It was computed assuming the beam to be Gaussian with full-width half maximum of $\theta_{FWHM} = 40\ \text{arcmin}$. The pixel window function corrects a suppression of power on small scales caused by the effect of the finite size of the {\tt HEALpix} pixel. For detailed discussion, we refer to the {\tt HEALpix} manual\footnote{\url{https://healpix.sourceforge.io/}}. 

Although Eq.~\ref{eq.pseudo_aps_estimate} provides an unbiased estimate for pure 21-cm maps, it becomes biased when applied to reconstructed maps. We employ a debiasing procedure \cite{fornazier2022bingo, marins2022foreground, de2023testing} to correct the estimated angular power spectrum (APS) for two effects: an additive bias from instrumental noise and a multiplicative bias arising from signal attenuation during the foreground removal process. Our debiased APS estimator is given by:
\begin{align}
    \hat{S}^{ij, \text{debiased}}_\ell =  \hat{S}^{ij, \text{biased}}_\ell / \beta_\ell  - \langle \hat{S}^{\text{noise}}_\ell \rangle .
    \label{eq.debiasing}
\end{align}
Here, $ \hat{S}^{ij, \text{biased}}_\ell $ is the biased APS estimate from a foreground-cleaned map, and $ \langle \hat{S}^{\text{noise}}_\ell \rangle $ is the ensemble average noise power spectrum estimated from random Gaussian realizations (see Sec.~\ref{sec.instrumental}). The factor $ \beta_\ell $ corrects for signal loss induced by the component separation pipeline. We estimate $ \beta_\ell $ as:
\begin{align}
    \beta_\ell = \bigg\langle \frac{\hat{S}^{ij, \text{biased}}_\ell}{\hat{S}^{ij, \text{Input}}_\ell} \bigg\rangle,
\end{align}
where $ \hat{S}^{ij, \text{Input}}_\ell $ is the APS of the input (21-cm + white noise) simulation. The average $ \langle \cdot \rangle $ is computed, when debiasing a given target simulation, from the remaining 2999 simulations to avoid statistical coupling. 

We further perform a bandwidth binning on the $S_\ell$s. This binning acts on the measurement in a way that decorrelates mixed modes. We bin the $\ell$ values into bins $\Delta \ell$ of width 11 (so e.g., the first bin is $2 \leq \ell \leq 12$ ). For each bin, we calculate a weighted average of the $\hat{S}_\ell$ (weighted by the number of spherical harmonic coefficients),
\begin{align}
    \hat{S}^{ij}_L = \frac{ \sum_{\ell = \ell_0}^{\ell_0 + \Delta \ell -1}(2\ell + 1)\hat{S}^{ij}_\ell }{ \sum_{\ell = \ell_0}^{\ell_0 + \Delta \ell -1} (2\ell + 1) } \,,
    \label{eq.bandwidth_binning}
\end{align}
where $\ell_0$ is the first multipole of each bin, e.g., $2, 13, 24, \cdots$. Here, $L$ simply denotes each multipole bin. 

Figure \ref{fig.relative_error} illustrates the relative error $(\hat{S}_L - S^{th}_L)/\sigma_L$, computed in relation to the {\tt UCLCl} theoretical $S^{th}_L$ at the fiducial cosmology. This plot draws attention to the behavior of the errors for a range of multipoles with respect to both the {\tt UCLCl} theoretical spectrum and the standard deviation of the 3000 simulations.

While the average $\hat{S}^{ij}_L$ values from the 3000 simulations closely track the {\tt UCLCl} theoretical spectrum, it is evident that the random errors on the spectrum for individual realizations, i.e. cosmic variance, exhibit strong correlations among neighboring multipoles within the same frequency channel. Let us take, for example, the frequency channel 10. The realization depicted in blue and in green shows mostly positive errors, while the realization depicted in red show mostly negative errors within the depicted multipole range.

Figure \ref{fig.relative_error} allows us to distinguish between random and systematic errors in the power spectrum recovery. The colored lines, representing individual simulation realizations, are dominated by random errors. The correlation of these errors across neighboring multipoles is introduced by the sky mask applied to the lognormal simulations. When averaged over all realizations, however, the black line is is consistent with zero across most of the multipoles, which shows that our pipeline recovers well the theoretical spectrum. The only notable deviation is a minimal systematic bias at very low multipoles ($\ell \lesssim 25$), which we attribute to the foreground removal process.

\section{Covariance Matrix}
\label{sec.covariance_matrix}

\begin{figure*}
    \centering
    \includegraphics[scale=0.5]{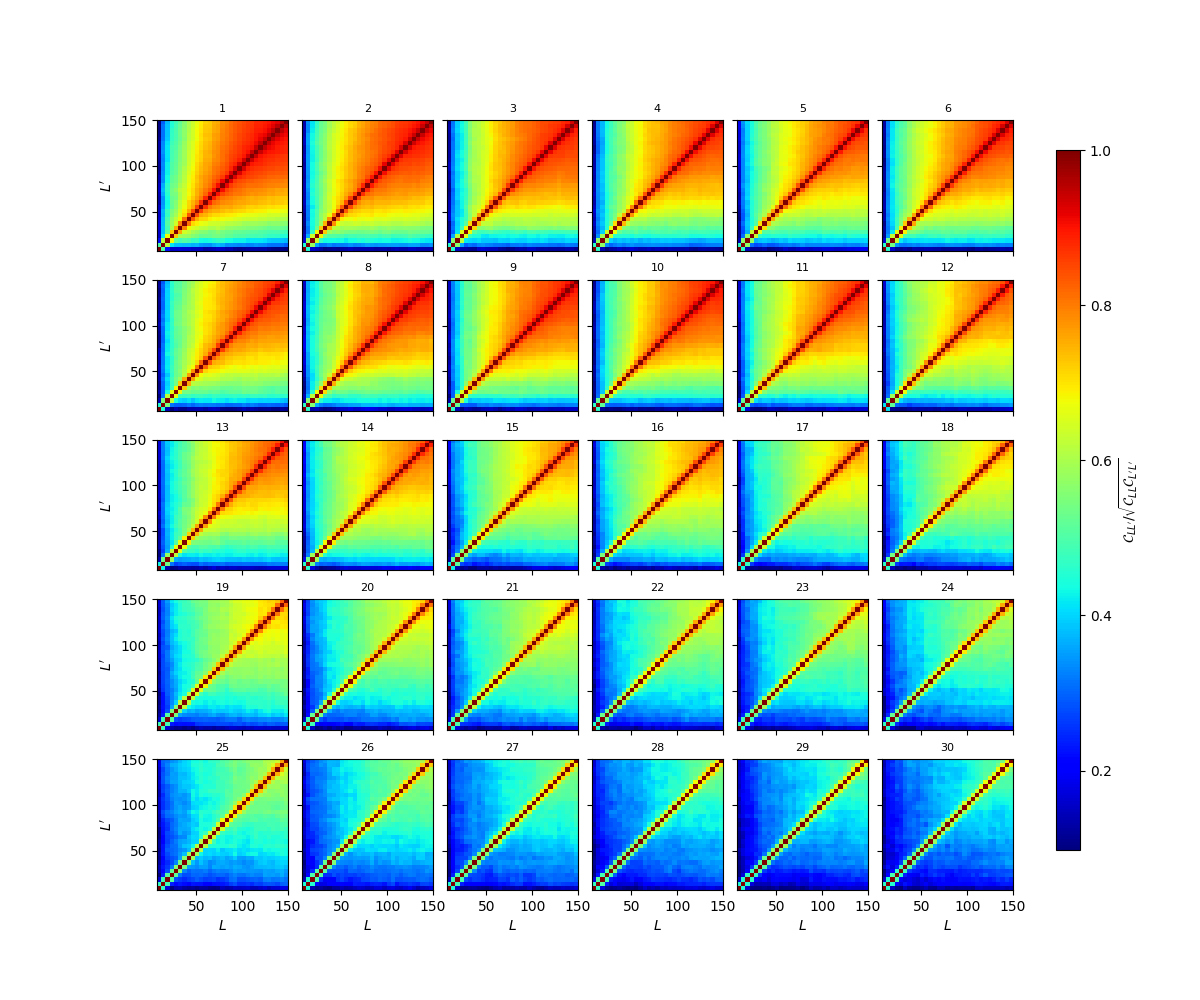}
    \caption{Pearson correlation coefficient $\mathcal{C}_{LL^\prime}/\sqrt{\mathcal{C}_{LL}\mathcal{C}_{L^\prime L^\prime}}$ where $\mathcal{C}_{LL^\prime}$ is the covariance matrix. In this plot, we did not include any systematics, so the correlation represents the pure 21-cm signal. There are two overall behaviors: the correlation increases with multipole, and the correlation decreases with redshift.}
    \label{fig.covariance_pearson_21cm}
\end{figure*}

\begin{figure*}
    \centering
    \includegraphics[scale=0.5]{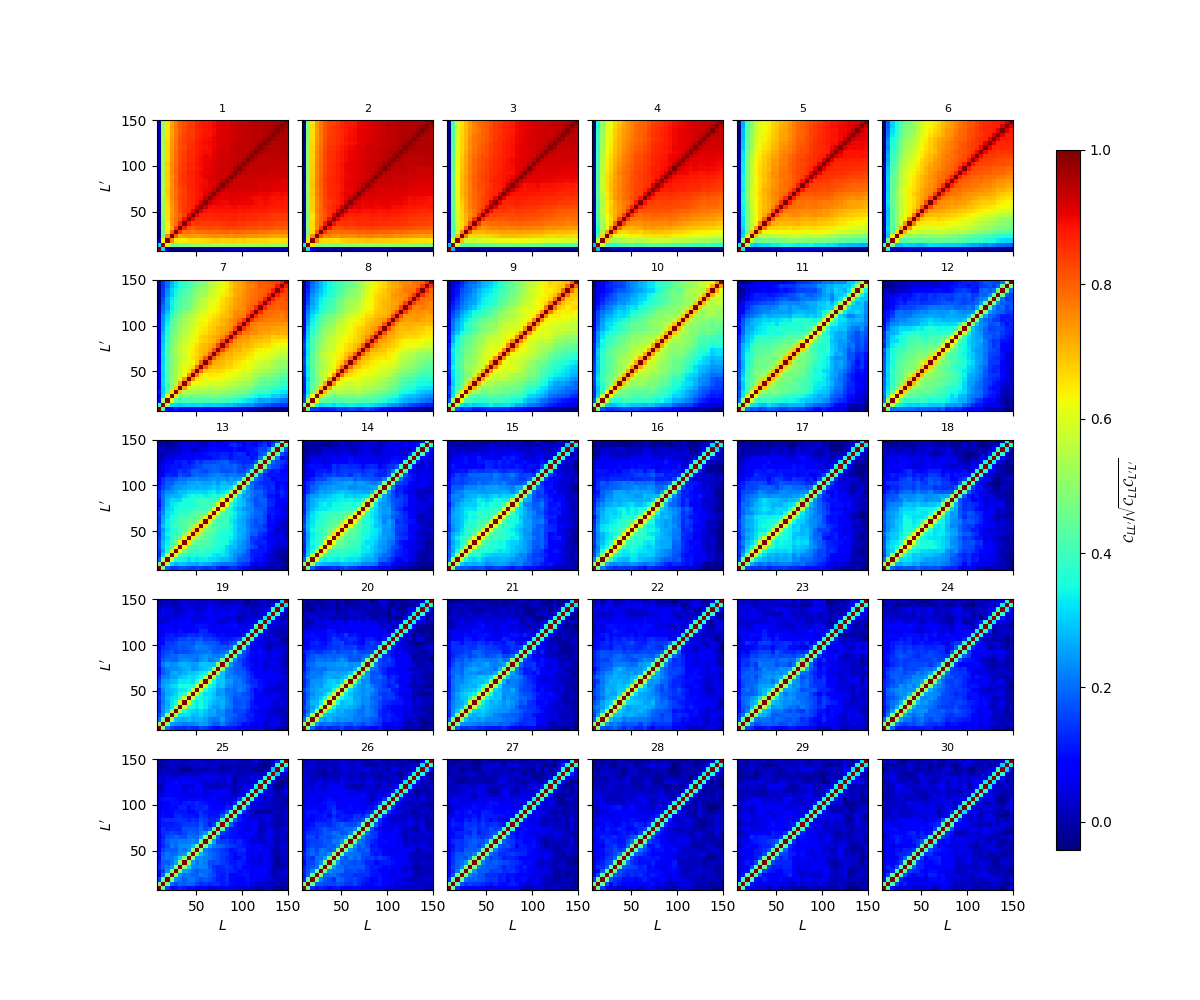}
    \caption{Pearson correlation coefficient $\mathcal{C}_{LL\prime}/\sqrt{\mathcal{C}_{LL}\mathcal{C}_{L\prime L\prime}}$ where $\mathcal{C}_{LL\prime}$ is the covariance matrix. White Nose and foreground removal systematic effects included. As in the pure 21-cm case, the correlation decreases with frequency channel. For the latest frequency channels, the correlation decreases for multipoles above $\ell \approx 100$ due to the deacrease of the signal-to-noise ratio. Additionaly for the earliest channels, the correlations remain high as an effect of foreground removal.}
    \label{fig.covariance_pearson_21cm_+_systematics}
\end{figure*}

For cosmological inference, it's essential to quantify the uncertainties associated with the observed power spectrum. To achieve this, we calculate the covariance matrix based on a set of $N_{sim} = 3000$ BINGO simulations, as elaborated upon in section \ref{sec.BINGO_sim}. Let $\hat{S}^{ij, n}_L$ be the weighted estimation of the angular power spectrum of the BINGO n-th realization, the covariance matrix of the ensemble of angular power spectra is 
\begin{align}
\mathcal{C}^{ij}_{LL'} = \frac{1}{N_{sim} -1} \sum_{n=1}^{N_{sim}} \left( \hat{S}^{ij,n}_L - \langle \hat{S}^{ij}_L\rangle \right) \left(  \hat{S}^{ij,n}_{L'} - \langle \hat{S}^{ij}_{L'}\rangle \right)^T \,,
    \label{eq.data_covariance}
\end{align}
where
$\langle \hat{S}^{ij}_L \rangle $ is the average pseudo-$C_\ell$ estimate for the 3000 BINGO simulations.
Figs. \ref{fig.covariance_pearson_21cm} and \ref{fig.covariance_pearson_21cm_+_systematics} show the Pearson correlation coefficient $\mathcal{C}_{LL'}/\sqrt{\mathcal{C}_{LL}\mathcal{C}_{L'L'}}$ computed from the {\tt FLASK} covariance matrix. The degree of correlation among different multipoles can be inferred from the diagonal properties of this matrix. The figures presented illustrate the correlation characteristics of both the pure simulated 21-cm cosmological signal and the debiased signal after accounting for systematic error sources.

In both instances, the correlation decreases with the redshift, leading to matrices with more pronounced diagonal features for the latest frequency channels. This affect is due to the stronger non-Gaussian features in lower redshifts. However, a distinction arises when comparing the pure 21-cm signal's correlation with that of the reconstructed signal incorporating systematic error sources.

The pure 21-cm signal covariance exhibits strengthening off-diagonal correlations at higher multipoles. This trend, driven by the non-Gaussian nature of the cosmological signal, is not observed in the debiased signal after white noise have been included. The reason for this divergence is the dominance of instrumental white noise at high multipoles, which follows Gaussian statistics. This substantial Gaussian noise component effectively dominates over the non-Gaussian correlations intrinsic to the 21-cm signal. Furthermore, the process of foreground reconstruction modifies the covariance structure, particularly in the earliest channels, where it introduces an increase in correlation. Thus, the final covariance of the reconstructed signal reflects these two effects combined.

\section{Likelihood and sampling methods}
\label{sec.cosmological_analysis}

The Bayesian framework has been widely used in cosmology to make inferences about the parameters of cosmological models in light of data from the large-scale structure (LSS). We follow a standard Bayesian analysis framework as commonly performed in the literature, e.g., \cite{krolewski2021cosmological, contarini2022euclid, loureiro2019cosmological, kohlinger2017kids, thomas2011angular, abbott2019dark, blake2007cosmological}. 

\subsection{Priors}
In this work, we take the standard approach of assuming that the priors are uniformly distributed (flat priors). Flat priors are used in order to reduce the dependence of the results on the choice of prior, allowing the
data to drive the results and avoid potential biases from strong prior beliefs \citep{massimi2021cosmic, diacoumis2019prior, efstathiou2008limitations}. The prior defined for the cosmological analysis are shown in table \ref{table.priors2}. The first block consists of six independent cosmological parameters: $\Omega_b h^2$, $\Omega_c h^2$, $100\theta_S$, $n_s$, $\ln 10^{10} A_s$, $\tau_r$, where $\Omega_b$ is the baryonic matter density, $\Omega_{c}$ is the cold dark matter density, $h$ is the Hubble constant, $A_s$ is the amplitude of the primordial power spectrum, $n_s$ is the spectral index, $\theta_s$ is the sound horizon at decoupling and $\tau_r$ is the optical depth at reionization epoch. The cosmological priors are defined to cover at least the $3\sigma$ interval of the Planck constraints from \cite{planck2016-l06}. The second block is the HI parameters, which are two per bin: the HI bias $b_{\rm HI}^i$ and the HI density ($\Omega_{\rm HI}^i$). The HI parameters are assumed constant within each frequency channel and scale independent. 

\begin{table}
\caption{Prior ranges for the cosmological analysis of BINGO + Planck 2018. This prior range was defined taking Planck constraints from \cite{planck2016-l06} as reference. We also included the prior of the HI parameters. The priors on $w_0$, $w_a$ were only used for $w_0 w_a$CDM constraints.}
\label{table.priors2}
\begin{center}
\begin{tabular}{ c | c  }
 \hline
 Parameter & Prior  \\ 
 \hline
 $\Omega_b h^2$ & 0.0219, 0.0228  \\  
 $\Omega_{c} h^2$ & 0.116, 0.124 \\
 $100 \theta_{s} $ & 1.03, 1.05 \\
 $n_s$ & 0.953, 0.978 \\
 $\ln 10^{10} A_s$ & 3.00, 3.17 \\
 $\tau_r$ & 0.0325, 0.763 \\ 
 \hline
 $ b_{\rm HI}^i $ & 0.5, 1.5 \\ 
  $ \Omega_{\rm HI}^i b_{\rm HI}^i $  &  $1 \times 10^{-4}$, $7 \times 10^{-4}$ \\
  \hline
 $w_0$ & -1.4,0 \\
 $w_a$ & -4,4 \\  
 \hline
\end{tabular}
\end{center}
\end{table}

\subsection{Multipole selection}

While the $C_\ell$ discussed in Section \ref{sec.theo_cl} incorporates dark matter non-linearities, it omits non-linear effects stemming from HI physics, such as scale-dependent bias. To address this, we opted to establish cuts in $\ell_{max}$ for each tomographic frequency channel, a strategy aimed at excluding non-linear scales. We computed the linear and non-linear $C_\ell$ at fiducial cosmology (the same used in Sec \ref{sec.cosmological_signal}) and performed a cut in $\ell_{max}$ where the percent deviation between the linear and non-linear models were smaller than 10\%. Additionally, we drop out small multipoles where foreground removal systematic errors appear. We fix $\ell_{min}=25$ for all multipoles. The employed multipole selection, along with the bandwidth binning, yields a relatively small number of data points for the initial frequency channels. Expanding the range of $\ell_{max}$ could enhance the precision of the constraints. However, given the absence of a non-linear theory for {\tt Hi} physics, our objective is to derive realistic constraints from the linear regime. The specific $\ell_{min}$ and $\ell_{max}$ values resulting from this approach for all frequency channels are presented in Table \ref{tab.multipole_selection}.

\begin{table}
\caption{Multipole range considered for cosmological analysis. The $\ell_{max}$ represents the multipole in which the percent deviation between the linear and non-linear models are smaller than 10\%. The $\ell_{min}$ was selected to exclude low multipoles with high systematic errors from the foreground removal. We also apply a bandwidth binning of $\Delta \ell = 11$. Each bin has a frequency width of 9.33MHz. The frequencies shown are the central frequency for each bin.}
\label{tab.multipole_selection}
\begin{center}
\begin{tabular}{ c | c | c | c  }
\hline
 Frequency channel & Freq. (MHz) & $\ell_{min}$ & $\ell_{max}$  \\ 
 \hline
  1 & 1255.33  & 25 & 42  \\
  2 & 1246.00  & 25 & 44  \\ 
  3 & 1236.67   & 25 & 48  \\ 
  4 & 1227.33  & 25 & 52  \\ 
  5 & 1218.00  & 25 & 54  \\
  6 & 1208.67  & 25 & 58  \\
  7 & 1199.33  & 25 & 61  \\
  8 & 1190.00  & 25 & 65  \\
  9 & 1180.67  & 25 & 69  \\
  10 & 1171.33 & 25 & 72  \\
  11 & 1162.00  & 25 & 77  \\
  12 & 1152.67  & 25 & 79  \\
  13 & 1143.33  & 25 & 85  \\
  14 & 1134.00 & 25 & 89  \\
  15 & 1124.67  & 25 & 93  \\
  16 & 1115.33  & 25 & 97  \\
  17 & 1106.00  & 25 & 100  \\
  18 & 1096.67  & 25 & 106 \\
  19 & 1087.33  & 25 & 111 \\
  20 & 1078.00  & 25 & 114 \\
  21 & 1068.67  & 25 & 119 \\
  22 & 1059.33  & 25 & 125 \\
  23 & 1050.00  & 25 & 129 \\
  24 & 1040.67  & 25 & 135  \\
  25 & 1031.33  & 25 & 139  \\
  26 & 1022.00 & 25 & 145  \\
  27 & 1012.67 & 25 & 150  \\
  28 & 1003.33 & 25 & 155  \\
  29 & 994.00 & 25 & 161 \\
  30 & 984.66  & 25 & 165  \\
  \hline
 \end{tabular}
\end{center}
\end{table}

\subsection{Likelihood}

The likelihood quantifies the agreement between the hypothesis and the data. Following \cite{loureiro2019cosmological}, we assume the likelihood to have the Gaussian form,
\begin{align}
    \mathcal{L} (\mathbf{\Theta}) &=  \frac{1}{\sqrt{\vert 2\pi \mathcal{C} \vert}} \exp\left( -\frac{1}{2} \chi^2 (\mathbf{\Theta}) \right) \,,
    \label{eq.likelihood_formula}
\end{align}
where
\begin{align}
    \chi^2 (\mathbf{\Theta}) &= \left[ \hat{S}_{L} - S^{th}_{L}(\mathbf{\Theta} )\right]^T 
 \mathcal{C}^{-1} \left[ \hat{S}_{L} - S^{th}_{L}(\mathbf{\Theta})\right] \,. \end{align}
The $\hat{S}_{L}$ is the pseudo-APS estimate measured from our dataset as described in Sec. \ref{sec.pseudo_aps_estimate}, $ S^{th}_{L}(\mathbf{\Theta})$ is the theoretical spectrum calculated with {\tt UCLCl} code following the Sec. \ref{sec.theo_cl} after being convolved with the mixing matrix (Eq. \ref{eq.PseudoAPS} ) and being binned with the same bandwidth of the data (Eq. \ref{eq.bandwidth_binning}). Finally, $\mathcal{C}$ is the covariance matrix described in Sec. \ref{sec.pseudo_aps_estimate} encapsulating the uncertainties and correlations between the data points. 

The presented likelihood function (Eq. \ref{eq.likelihood_formula}) is based on the assumption that the observed data follows a Gaussian distribution. However, this approximation is imperfect due to the lognormal field's inherent non-Gaussian characteristics. To explore the potential of non-Gaussian likelihoods, further investigation and testing could be pursued in future endeavors.

Tighter constraints on the cosmological parameters can be obtained by combining BINGO with the Planck 2018 dataset. Assuming that each probe provides independent information, the joint likelihood is
\begin{align}
   \mathcal{L}_{\,\text{BINGO + Planck}}  
 = \mathcal{L}_{\,\text{BINGO}} \times \mathcal{L}_{\,\text{Planck}} \,.
\end{align}
The CMB data from Planck was added through the Planck likelihood codes \texttt{Commander} and \texttt{Plik} \citep{planck2016-l05}. We include in our analysis the likelihoods lowlTT (temperature data over $2 \leq \ell \leq 30$), Planck TT,TE,EE (the combination of Planck TT, Planck TE, and Planck EE, taking into account correlations between the TT, TE, and EE spectra at $\ell > 29$), lowE (the EE power spectrum over $2 \leq \ell \leq 30$) and the Planck CMB lensing likelihood. The multiplication of these four likelihoods is commonly referred as  Planck TT,TE,EE+lowE+lensing.

\subsection{Sampling}

Sampling points in parameter space play a crucial role in Bayesian inference, as they allow us to estimate the posterior distribution of the parameters. Some traditional MCMC methods are Metropolis–Hastings \citep{metropolis1953equation, hastings1970monte}, Gibbs sampling \citep{geman1984stochastic}, Hamiltonian sampling \citep{duane1987hybrid}, and thermodynamic
integration \citep{ruanaidh2012numerical}. In our work, we use the nested sampling method \citep{skilling2004nested, skilling2006nested}. It is more efficient than traditional MCMC methods because it uses the information about the prior volume to guide the sampling instead of randomly proposing moves in the parameter space as in Metropolis–Hastings. The algorithm works by constructing a sequence of nested sets of points in parameter space, each set being a subset of the previous one, and keeping track of the points that have the highest likelihood. {\tt Multinest} \citep{feroz2008multimodal, feroz2009multinest, feroz2013importance} is the most popular and freely accessible software package that implements the nested sampling method. In this paper, we employ {\tt Pliny} \citep{rollins2015chemical}, an open-source nested sampling algorithm implementation that prioritizes optimal parallel performance on both distributed and shared memory computing clusters. {\tt Pliny} has been successfully applied to cosmological analysis as in \cite{mcleod2017joint, loureiro2019upper, loureiro2019cosmological, jeffrey2019parameter}. 

\section{Results}
\label{sec.results}


\begin{table*}
\centering
\caption{Marginalized cosmological $\Lambda$CDM constraints and 68\% credible intervals for Planck 2018 and BINGO + Planck 2018. The first column shows Planck-only constraints. The next three columns correspond to BINGO + Planck using three different realizations (1, 2, 3) of the pure 21-cm cosmological signal. The last two columns show BINGO + Planck constraints for realization 1 including white noise (WN) and foreground removal, respectively.}
\label{tab.bingo_planck_constraints}
\renewcommand{\arraystretch}{1.3}
\begin{tabular}{@{} l *{6}{c} @{}}
\toprule
Parameter & 
  Planck &
  \makecell{Planck + BINGO \\ (realization 1)} & 
  \makecell{Planck + BINGO \\ (realization 2)} & 
  \makecell{Planck + BINGO \\ (realization 3)} &
  \makecell{Planck + BINGO \\ (21cm + WN)} & 
  \makecell{Planck + BINGO \\ (Reconstructed)} \\
\midrule

$\Omega_{b} h^2$ 
  & $0.02240 \pm 0.00017$ 
  & $0.022319^{+0.000075}_{-0.000079}$ 
  & $0.022335^{+0.000066}_{-0.000064}$ 
  & $0.022290^{+0.000060}_{-0.000059}$ 
  & $0.022336^{+0.000075}_{-0.000074}$ 
  & $0.022309 \pm 0.000072$ \\

$\Omega_{c} h^2$ 
  & $0.1188^{+0.0014}_{-0.0013}$ 
  & $0.11984^{+0.00063}_{-0.00065}$ 
  & $0.11977^{+0.00050}_{-0.00049}$ 
  & $0.12028^{+0.00051}_{-0.00049}$ 
  & $0.11947^{+0.00058}_{-0.00060}$ 
  & $0.12023^{+0.00053}_{-0.00052}$ \\

$H_0$ 
  & $67.69 \pm 0.63$ 
  & $67.23 \pm 0.29$ 
  & $67.25^{+0.22}_{-0.21}$ 
  & $67.04 \pm 0.21$ 
  & $67.39^{+0.27}_{-0.26}$ 
  & $67.06 \pm 0.23$ \\

$\ln 10^{10}A_s$ 
  & $3.076^{+0.029}_{-0.027}$ 
  & $3.0548^{+0.0086}_{-0.0095}$ 
  & $3.0517^{+0.0088}_{-0.0087}$ 
  & $3.0483^{+0.0084}_{-0.0081}$ 
  & $3.0487^{+0.0095}_{-0.0096}$ 
  & $3.0528^{+0.0080}_{-0.0084}$ \\

$n_s$ 
  & $0.9672^{+0.0046}_{-0.0045}$ 
  & $0.9655^{+0.0020}_{-0.0022}$ 
  & $0.9665 \pm 0.0016$ 
  & $0.9633 \pm 0.0019$ 
  & $0.9665 \pm 0.0020$ 
  & $0.9654^{+0.0016}_{-0.0017}$ \\

$\tau_{r}$ 
  & $0.0502^{+0.0077}_{-0.0079}$ 
  & $0.0477^{+0.0033}_{-0.0035}$ 
  & $0.0498 \pm 0.0030$ 
  & $0.0509 \pm 0.0028$ 
  & $0.0467^{+0.0031}_{-0.0028}$ 
  & $0.0482 \pm 0.0031$ \\

\bottomrule
\end{tabular}
\end{table*}


In this section, we present cosmological constraints for the BINGO simulations + Planck 2018 likelihood within the framework of $\Lambda$CDM and $w_0w_a$CDM cosmology. In subsection \ref{sec.bingo_planck_reali}, we study the consistency of parameter constraints across different realizations of the pure 21-cm cosmological signal BINGO simulation within $\Lambda$CDM cosmology. In subsection \ref{sec.bingo_planck_sys} we explore the impact of white noise and residual of foreground removing on the cosmological parameter contours. In subsection \ref{sec.nuisance_constraints}, we discuss the constraints on the HI parameters, and in subsection \ref{sec.w0wa} we discuss the constraints for $w_0w_a$CDM cosmology parameters. Finally, in subsection \ref{sec.goodness_fits}, we discuss the goodness of the Monte Carlo fits.

\subsection{Cross-checking $\Lambda$CDM constraints between different 21-cm realizations}
\label{sec.bingo_planck_reali}

\begin{figure*}
    \centering
    \includegraphics[scale=0.5]{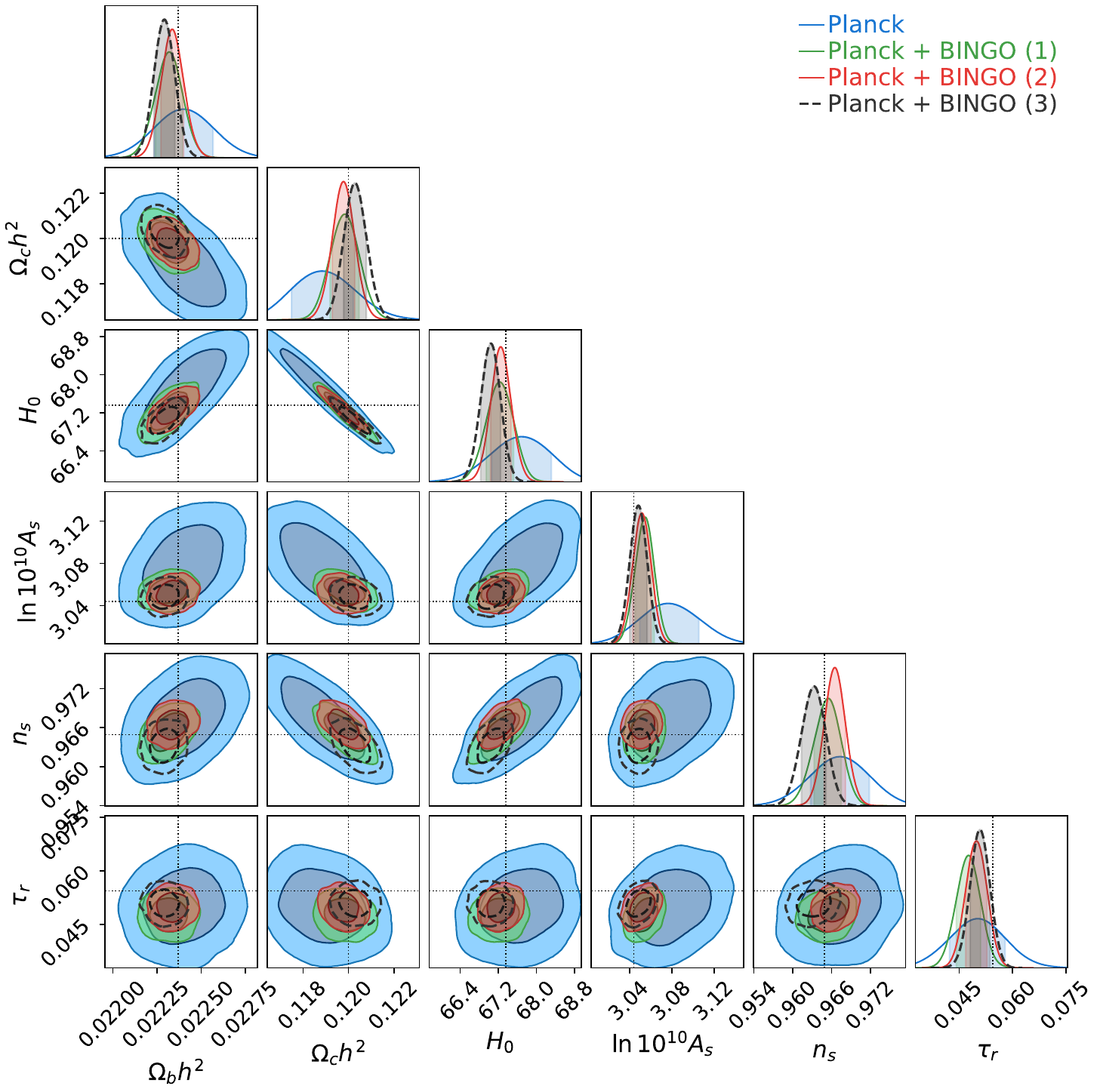}
    \caption{The blue contour represents the posterior from Planck 2018 data while the others BINGO + Planck 2018 $\Lambda$CDM contours. The green, red and black dashed contours correspond to different BINGO realizations. The dotted black lines indicate the truth (fiducial) value of the parameters. It does not include the systematic effects of White Noise and foreground removal. We find consistent results across the different realizations.}
    \label{fig.BINGO_Planck_compare_realizations}
\end{figure*}

We derived constraints on cosmological parameters by jointly analyzing simulated data from BINGO and the \textit{Planck} 2018 likelihood under the assumption of a flat $\Lambda$CDM framework. \textit{Planck}'s measurements provide more stringent constraints than other independent cosmological experiments, making them a robust baseline for comparison. The analysis focused on six primary cosmological parameters: the baryon density $\Omega_b h^2$, cold dark matter density $\Omega_c h^2$, angular acoustic scale $100\theta_s$, scalar spectral index $n_s$, amplitude of primordial fluctuations $\ln 10^{10} A_s$, and optical depth to reionization $\tau_r$, alongside 60 additional parameters characterizing neutral hydrogen (HI). The Hubble constant ($H_0$) was inferred as a derived parameter from the aforementioned six. We compared results using the pseudo-$C_\ell$ estimator across multiple realizations of the BINGO simulations, with the outcomes for three realizations
illustrated in Fig.\ref{fig.BINGO_Planck_compare_realizations}. See also Table \ref{tab.bingo_planck_constraints}.

Initially, we examined the parameter fitting outcomes under idealized conditions, considering only the pure cosmological 21-cm signal. While this analysis excluded observational systematics such as instrumental noise and foreground contamination, it consistently accounted for the sky mask and the beam convolution. Our findings demonstrate that combining BINGO simulations with Planck 2018 data significantly improves the precision of parameter constraints compared to Planck 2018 alone. The joint analysis yields tighter posteriors that remain fully consistent with the Planck 2018 fiducial cosmology.

The joint analysis significantly tightens the constraint on the optical depth to reionization ($\tau_{reio}$), despite the 21 cm power spectrum having no direct sensitivity to this parameter. The inclusion of BINGO simulations provides independent measurements of $A_s$ and $n_s$. By reducing uncertainties in these primordial parameters, the joint analysis indirectly diminishes the allowed range for $\tau_{reio}$.

Notably, the probability contours exhibit slight positional shifts between different BINGO realizations. These variations arise from statistical fluctuations inherent to the simulated data and theoretical degeneracies in the parameter space. Crucially, the magnitude of these shifts aligns with the expected statistical uncertainties at the 68\% and 95\% confidence levels. Despite these minor displacements, the contours retain consistent shapes and sizes across all realizations, indicating robustness in the parameter estimation methodology.

\subsection{Exploring $\Lambda$CDM constraints with white noise and foreground removing effects included}
\label{sec.bingo_planck_sys}

\begin{figure*}
    \centering
    \includegraphics[scale=0.5]{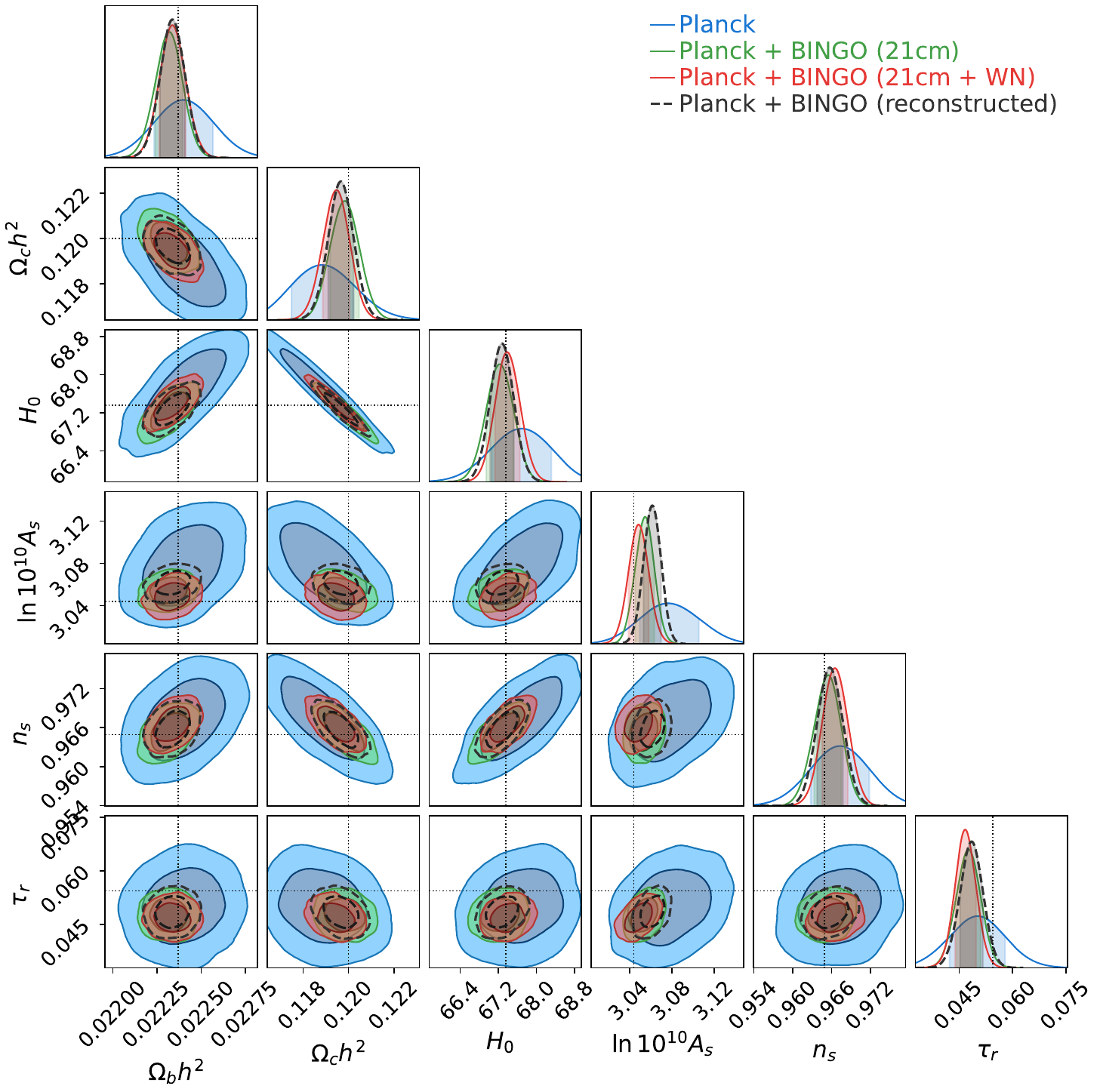}
    \caption{The blue contour represents the posterior from Planck 2018 data while the others BINGO + Planck 2018 $\Lambda$CDM contours. The green contour were computed for a BINGO simulation that only included the cosmological 21-cm signal, the red contour assumed the same 21-cm realization but included white noise, and the black dashed contour also included foreground removal. The dotted black lines indicate the truth (fiducial) value of the parameters.}
    \label{fig.BINGO_Planck_countours_systematics}
\end{figure*}

Initially, we examined the parameter fitting outcomes with sole consideration for the pure 21-cm cosmological signal. Although the previews section excludes observational systematics such as instrumental noise and foregrounds, the results always take into account the effects of the sky mask and beam convolution. We now extend our analysis to include the impact of white noise and foreground removal. To achieve this, both the measured $C_\ell$ values, designated for fitting, and the associated covariance matrix were sourced from BINGO simulations that incorporated white noise, as elaborated upon in Section \ref{sec.BINGO_sim}. Finally, we took into account the residual foregrounds following the methodology outlined in Section \ref{sec.BINGO_sim}. The constraints are shown in Table \ref{tab.bingo_planck_constraints} and Fig. \ref{fig.BINGO_Planck_countours_systematics}.

Notably, the constrained values of the $\Lambda$CDM parameters exhibited consistent behavior across the three scenarios. The confidence levels associated with these estimates remained similar across the different cases. This shows us that systematic error sources do not increase the uncertainty surrounding our parameter estimates. The fact that the uncertainties do not increase can be explained by the high signal-to-noise ratio (SNR) within the multipole range used for the cosmological analysis and the low amplitude of the foreground removal residuals.

Our analysis underscores that incorporating BINGO data significantly refines the precision of parameter constraints in comparison to relying solely on the Planck likelihood, even when including white noise and foreground removal. For the simulation that includes white noise and foreground removal, we got $\Omega_b h^2 = 0.022309 \pm 0.000072$, $\Omega_{c} h^2 = 0.12023^{+0.00053}{-0.00052}$, $H_0 = 67.06 \pm 0.23$, $n_s = 0.9654^{+0.0016}{-0.0017}$, $\ln 10^{10}A_s = 3.0528^{+0.0080}{-0.0084}$, $\tau{r} = 0.0482 \pm 0.0031$. This corresponds to a decrease in the confidence level with respect to the Planck-only one by 58\% for $\Omega_b h^2$, 61\% for $\Omega_{c} h^2$ and 63\% for $H_0$.

\subsection{HI parameter constraints}
\label{sec.nuisance_constraints}

\begin{figure*}
    \centering
     \begin{subfigure}[b]{0.49\textwidth}
         \centering
         \includegraphics[width=\textwidth]{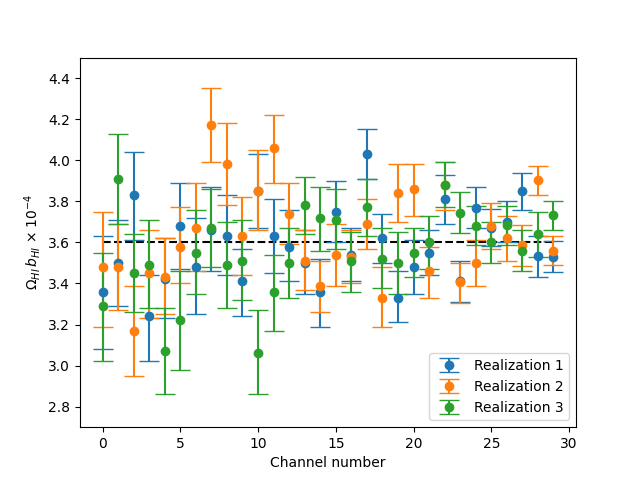}
         \caption{Constraints on $\Omega_{HI} b_{HI}$ for three different realizations.}
         \label{fig.Ohibhi_realization}
     \end{subfigure}
     \begin{subfigure}[b]{0.49\textwidth}
         \centering
         \includegraphics[width=\textwidth]{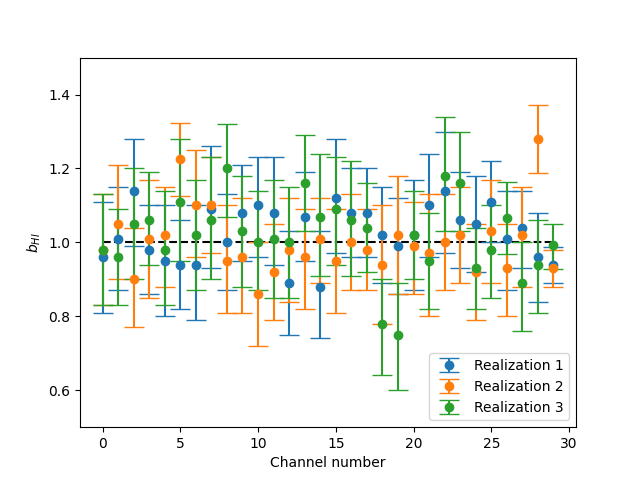}
         \caption{Constraints on $ b_{HI}$ for three different realizations.}
         \label{fig.bhi_realization}
     \end{subfigure}     
     \begin{subfigure}[b]{0.49\textwidth}
         \centering
         \includegraphics[width=\textwidth]{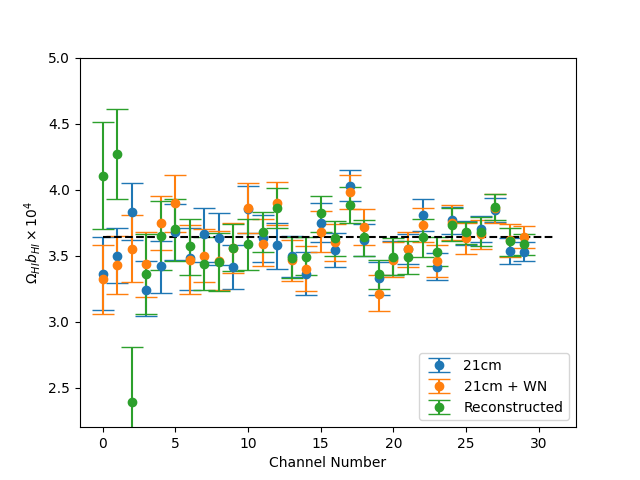}
         \caption{ Constraints on $ \Omega_{HI} b_{HI}$ for white noise and FG removing effects included.}
         \label{fig.Ohibhi_sys}
     \end{subfigure} 
     \begin{subfigure}[b]{0.49\textwidth}
         \centering
         \includegraphics[width=\textwidth]{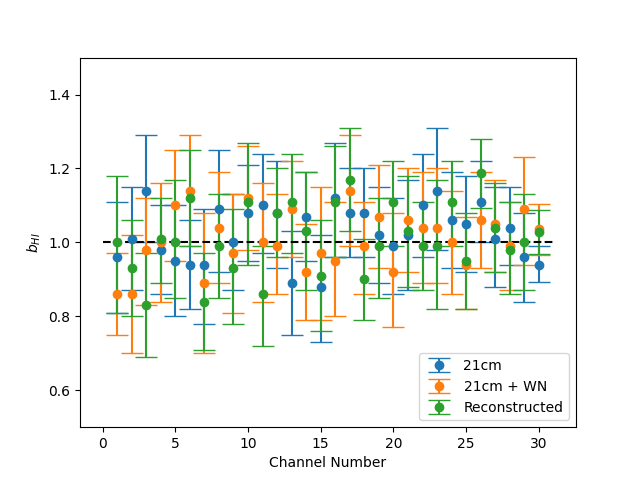}
         \caption{ Constraints on $ b_{HI}$ for white noise and FG removing effects included. }
         \label{fig.bhi_sys}
     \end{subfigure}  
     \caption{Marginalized constraints for the 60 HI parameters and 68\% credible intervals for BINGO + Planck likelihood. The dashed black lines indicate the truth (fiducial) value of the parameters. In figure \ref{fig.Ohibhi_realization} and \ref{fig.bhi_realization}, we show the constraints for the $\Omega_{HI}b_{HI} $ and $b_{HI}$ for three different realizations of the pure cosmological signal. In figure \ref{fig.Ohibhi_sys} and \ref{fig.bhi_sys} we show the constraints for three cases: simulation of the pure cosmological 21-cm signal (blue), simulation including white noise (orange), simulation including white noise and foreground removal (green). We find that we can recover the fiducial values within $1 \sim 2 \sigma$ for most of the parameters. This result shows the potential of BINGO to constrain HI history.}
     \label{fig.HI_1d}
\end{figure*}

\begin{figure*}
    \centering
     \begin{subfigure}[c]{0.48\textwidth}
         \centering
         \includegraphics[width=\textwidth]{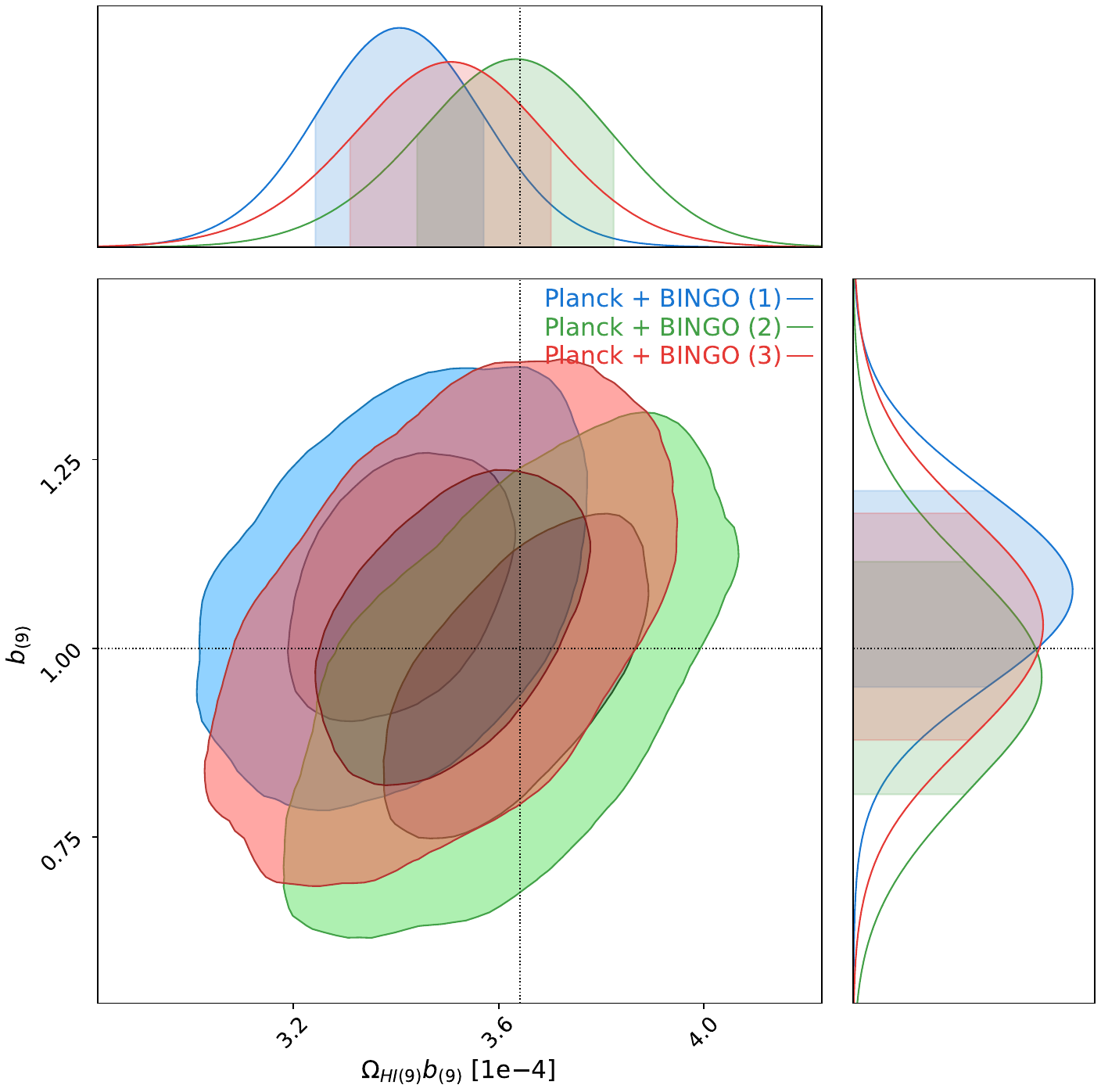}
         \caption{HI constraints for different realizations}
         \label{fig.HI_reali}
     \end{subfigure}
    \centering
     \begin{subfigure}[c]{0.48\textwidth}
         \centering
         \includegraphics[width=\textwidth]{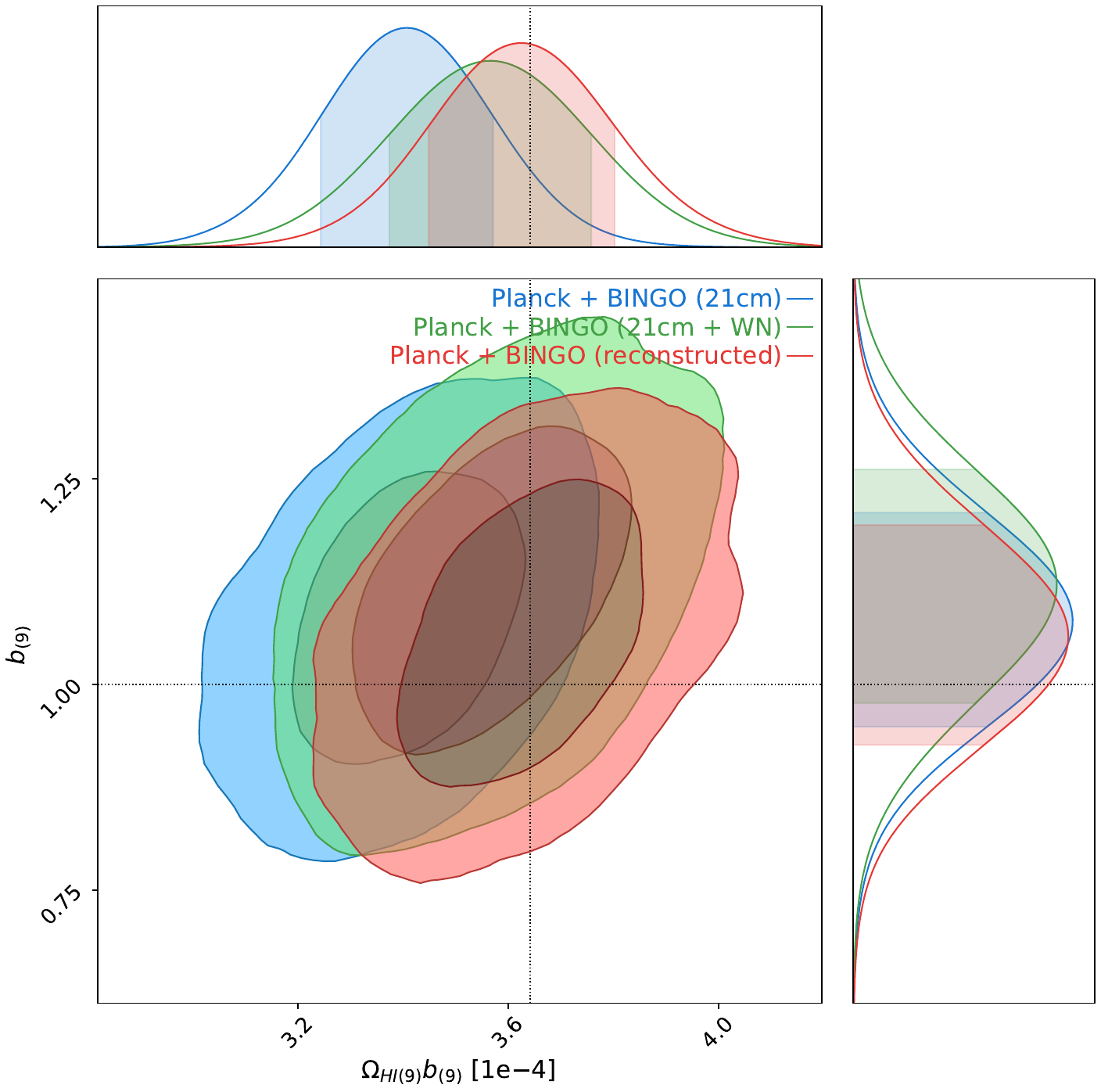}
         \caption{HI constraints for white noise and FG removing effects included}
         \label{fig.HI_sys}
     \end{subfigure}    
    \caption{2D constraints of the HI parameters of the 10th frequency channel of the Planck + BINGO likelihood. On the left we show the consistency across different realizations of of the BINGO simulations. The dotted black lines indicate the truth (fiducial) value of the parameters. On the right we compare the contours for the simulation of the pure 21-cm cosmological signal (blue), simulation including white noise (green), simulation including white noise and foreground removal (red). We find similar shapes and uncertainty levels across the comparisons.}
    \label{fig.HI_2d}  
\end{figure*}     

In this subsection, we discuss the constraints on the HI parameters for the simulations shown in the preview sections. In all of our analyzes, the parameters $\Omega_{\rm HI} b_{\rm HI}$ and $b_{\rm HI}$ were treated as independent variables across all frequency channels, resulting in a total of 60 distinct HI parameters.  The marginalized constraints for all 60 HI parameters with their respective 68\% credible interval are displayed in Fig. \ref{fig.HI_1d}. Aditionally, the 2D constraints on these parameters for one of the frequency channels are shown in Figs. \ref{fig.HI_2d}. Both figures shows consistencies of the marginalized values and uncertainty levels across different scenarios, and the potential of BINGO to measure the HI history.

Figs. \ref{fig.Ohibhi_realization} and \ref{fig.bhi_realization} presents the derived constraints on the HI parameters across three independent realizations of the 21-cm signal. The agreement between realizations shows that our methodology is stable against statistical variations in the input signal, with all chains converging to consistent posterior distributions.

To quantify the impact of observational effects, Figs. \ref{fig.Ohibhi_sys} and \ref{fig.bhi_sys} compare constraints under three scenarios:
pure cosmological signal without white noise or foreground contamination;
simulation including white noise; and simulation combining effects of white noise and residual of foreground removing. The fiducial input values (indicated by dashed lines) are recovered within $1\sim 2\sigma$ confidence intervals for all parameters, demonstrating the statistical consistency of our pipeline. Similarly as for the cosmological parameters, the uncertainties surrounding the parameter estimates do not increase with the addition of systematic error sources. Again, it can be explained by the high SNR and low residual level within the multipole range of the fitting.  

\subsection{ $w_0 w_a $CDM constraints }
\label{sec.w0wa}

\begin{figure}
    \centering    \includegraphics[width=0.48\textwidth]{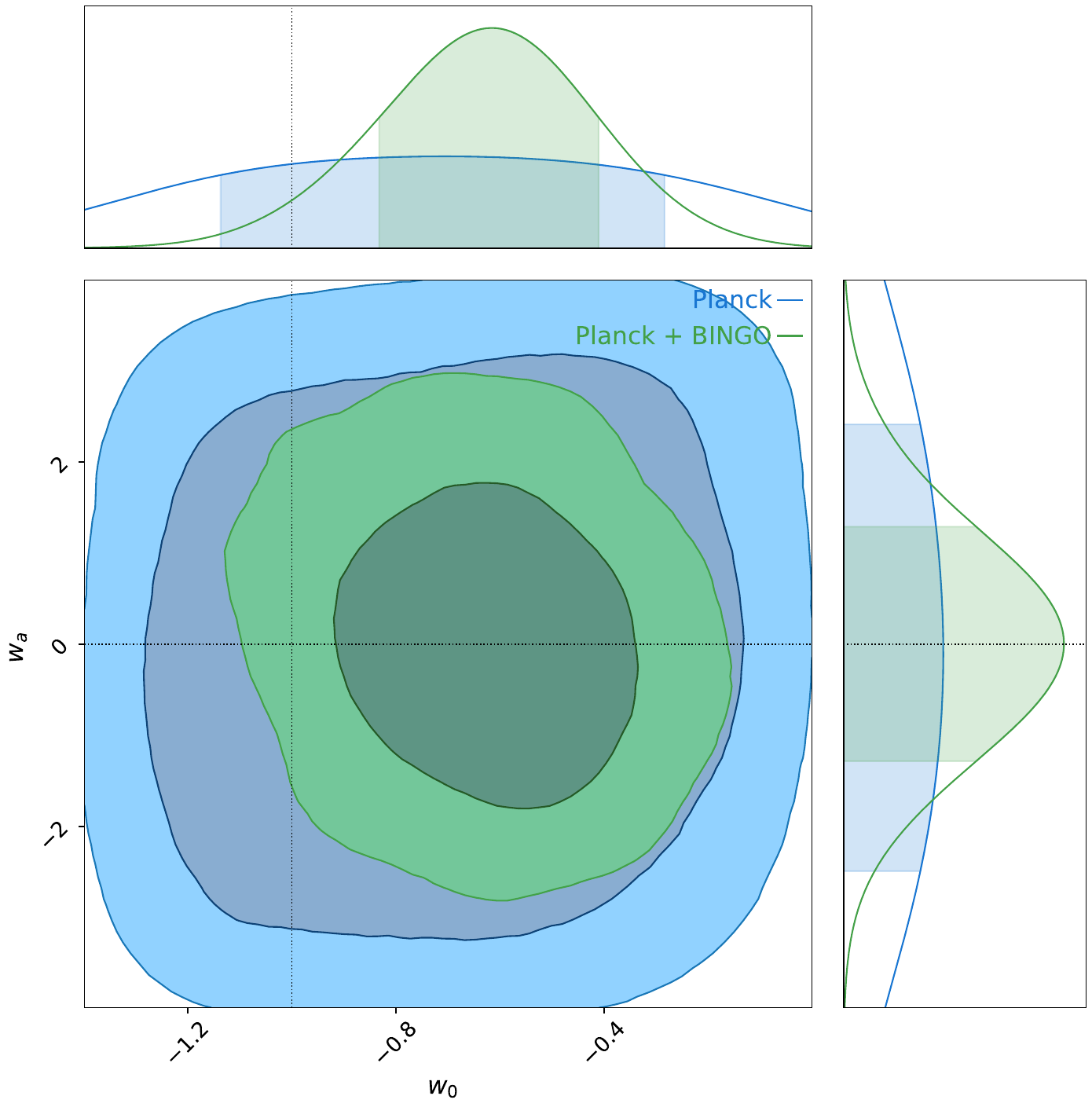}
    \caption{Constraints for $w_0w_a$CDM cosmology for Planck 2018 (blue) and Planck 2018 + BINGO (green). With the inclusion of the BINGO simulation data we can measure $w_0$ and $w_a$ with a confidence level of 0.22 and 1.2.}
    \label{fig.w0wa}
\end{figure}

The constraints on the $w_0w_a$CDM cosmological parameters, derived
from Planck 2018 data alone and in combination with the BINGO
simulation, are presented in Fig. \ref{fig.w0wa}. The $w_0w_a$CDM model parameterizes the dark energy equation of state as $w(z)=w_0+w_a (1-a)$ \citep{chevallier2001accelerating,linder2003exploring}, where $w_0$ is its present-day value and $w_a$ quantifies redshift evolution. Our analysis reveals a marked improvement in the precision of these parameters
when BINGO’s 21-cm intensity mapping data is incorporated into
the Planck baseline.

Planck 2018 alone (blue contours in Fig. \ref{fig.w0wa}), the posterior distributions of $w_0$ and $w_a$ largely reflect the prior volume, consistent with the known challenge of constraining dark energy parameters using CMB data alone \cite{planck2016-l06}. In contrast, combining Planck with BINGO’s simulated 21-cm observations (green contours) breaks this degeneracy, yielding a measurement of the dark
energy equation of state consistent with the fiducial $\Lambda$CDM values ($w_0=-1$, $w_a=0$) at the $1.9\sigma$ confidence level. The joint constraints are:
\begin{align}
w_0 =  -0.62^{+0.20}_{-0.22},  \quad
w_a =   0.0\pm 1.3 ,
\end{align}
This demonstrates BINGO’s potential to probe the nature of dark
energy. Although Planck yields precise measurements for most of the
cosmological parameters, BINGO complements Planck by anchoring
the low-redshift expansion history, where dark energy dominates. 

\subsection{Goodness of the fits}
\label{sec.goodness_fits}
\begin{figure*}
    \centering
\includegraphics[scale=0.35]{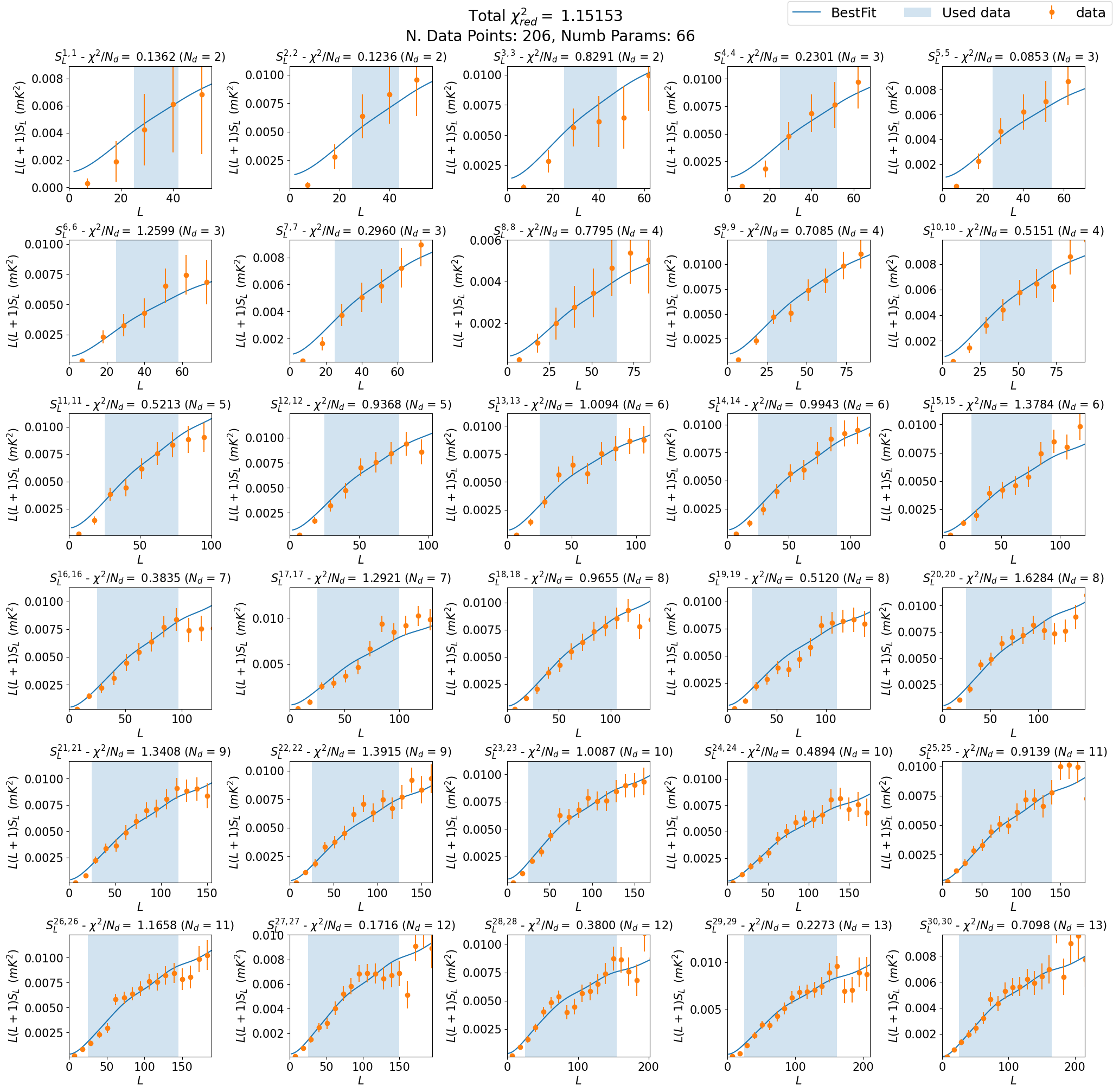}
    \caption{The blue solid curve shows the best-fit theoretical angular power spectrum ($S_L$) derived from the optimal cosmological and hydrogen intensity (HI) parameters. Orange points represent simulated BINGO data, including contributions from white noise and foreground removal. Error bars indicate the $1\sigma$ uncertainty on $S_L$, computed as the standard deviation across 3000 mock realizations. For each frequency channel, the number of data points ($N_d$) and the reduced $\chi^2$ statistic ($\chi^2/N_d$) are provided to quantify the agreement between theory and simulation.} 
\label{fig.best_fit_curve}
\end{figure*}

\begin{table}
\centering
\caption{Overall reduced $\chi^2$ for different BINGO realizations. The corresponding contour plots is Fig. \ref{fig.BINGO_Planck_compare_realizations}. It do not include white noise and foreground removing systematic effects. }
\label{tab.chi2_reali}
\begin{tabular}{ c|c|c } 
 \hline
Realization 1 & Realization 2 & Realization 3  \\
 \hline
 1.32 & 1.13 & 1.21  \\ 
 \hline
\end{tabular}
\centering
\caption{Overall reduced $\chi^2$ for the pure 21-cm signal scenario, 21-cm signal added with white noise and 21-cm signal added with white noise and foreground removal systematic effects included. The corresponding contour plots is Fig. \ref{fig.BINGO_Planck_countours_systematics}.}
\label{tab.chi2_sys}
\begin{tabular}{ c|c|c } 
 \hline
 21cm & 21cm + WN & Reconstructed \\
 \hline
 1.13 & 1.15 & 1.15 \\ 
 \hline
\end{tabular}
\centering
\caption{Overall reduced $\chi^2$ for the BINGO simulation with white noise and foreground removal systematic effects included. We compare two cosmological models, $\Lambda$CDM and $w_0w_a$CDM. }
\label{tab.chi2_cosmo}
\begin{tabular}{ c|c } 
 \hline
 $\Lambda$CDM & $w_0w_a$CDM \\
 \hline
 1.15 & 1.19 \\ 
 \hline
\end{tabular}
\end{table}

We calculated the overall reduced $\chi^2$ for the Monte Carlo fits presented in this paper. The overall reduced $\chi^2$ is defined as $\chi^2_{red} = \chi^2/(N_d - N_p)$, where $N_d$ is the number of data points and $N_p$ is the number of parameters. Given the chosen bandwidth binning and multipole selection for our analysis, the total number of data points is  $N_d = 206$. The number of parameters, $N_p$ is 66 (6 cosmological + 60 HI parameters) for the $\Lambda$CDM simulations. When $w_0$ and $w_a$ parameters are included, the number of parameters is 68. Fig. \ref{fig.best_fit_curve} visualizes the best-fit theoretical curve and corresponding data points.

Table \ref{tab.chi2_reali} showcases the $\chi^2_{red}$ values for different realizations of the BINGO simulation. These values span from 1.13 to 1.32 showing they have overall good fits with $\chi^2_{red}$ values close to one. 

Table \ref{tab.chi2_sys} provides $\chi^2_{red}$ values for three different scenarios: pure 21-cm signal, 21-cm signal with added random white noise, and 21-cm signal with both random white noise and systematic effects from foreground removal. The inclusion of white noise and foreground removal systematic effects leads to a minimal increase in $\chi^2_{red}$ compared to the parameter fitting using only the pure 21-cm signal. The $w_0w_a$CDM fitting also led to a minimal increase in $\chi^2_{red}$ (Table \ref{tab.chi2_cosmo}), although a decrease in the total $\chi^2$. All this cases suggest consistency in the quality of the fitting across the different scenarios of error sources and cosmological models.

\section{Conclusions}
\label{sec.conclusions}

In this study, we analyzed the potential of the BINGO 21-cm intensity mapping experiment, combined with \textit{Planck} 2018 CMB data, to constrain cosmological parameters. For the $\Lambda$CDM model, we jointly constrained six core parameters ($\Omega_b h^2$, $\Omega_c h^2$, $100\theta_s$, $n_s$, $\ln 10^{10} A_s$, and $\tau_{\mathrm{reio}}$) alongside 60 HI parameters and extended this framework to the $w_0w_a$CDM model by including dark energy parameters $w_0$ and $w_a$. Combining BINGO simulations with \textit{Planck} data tightens cosmological parameter constraints to $\sim$40\% of the size of those from \textit{Planck} alone (a $60\%$ reduction), with posteriors fully consistent with the fiducial \textit{Planck} cosmology.

We compared cosmological constraints across different BINGO simulation realizations and found only minor variations between them. The probability contours exhibit small displacements that align within the expected $68\%$ and $95\%$ confidence intervals, consistent with statistical uncertainties inherent to the simulated data. These fluctuations arise from both statistical noise in the simulations and theoretical degeneracies in the parameter space, underscoring the robustness of our methodology while highlighting the stability of the results across realizations.

The inclusion of white noise and foreground removal in our analysis revealed the robustness of the BINGO–\textit{Planck} joint constraints. While white noise introduces random fluctuations mimicking instrumental uncertainties, and foreground removal adresses systematic biases from imperfect separation of galactic/intergalactic signals, our parameter estimates remained stable. We found similar confidence levels across the studied scenarios: the pure cosmological 21-cm signal, the cosmological 21-cm signal with white noise added, and the reconstructed signal after full foreground removal (which includes white noise). This resilience is explained by BINGO’s high signal-to-noise ratio (SNR) within the multipole range ($\ell < 165$) used for cosmological inference, where the 21-cm signal dominates over noise. 

This forecast is built upon a pipeline that implements full, blind foreground removal. The constraints presented are consequently robust against this systematic. However, several simplifying assumptions remain in our simulation framework, which the reader should consider when assessing the confidence level of our results. These include the use of a Gaussian beam approximation, spectrally smooth foreground components, stationary white noise, and lognormal HI maps without non-linear clustering. These necessary and standard choices for a large-scale forecasting exercise define an optimistic but well-controlled benchmark. They successfully demonstrate the fundamental statistical power of BINGO and the efficacy of our debiasing method, while clearly identifying where future work must incorporate more complex instrumental and astrophysical systematics to transition to observational analysis.

Our analysis demonstrates BINGO’s capability to reconstruct the evolution of neutral hydrogen (HI) across cosmic time through measurements of $\Omega_{\mathrm{HI}} b_{\mathrm{HI}}$ and $b_{\mathrm{HI}}$ over 30 independent frequency channels. We found consistent parameter constraints across multiple realizations and observational scenarios, even in the presence of realistic noise and residual foreground contamination. Fiducial values were recovered within $1$–$2\sigma$ confidence intervals across frequency channels. However, we note that the quality of foreground reconstruction can affect parameter recovery for $\Omega_{\mathrm{HI}} b_{\mathrm{HI}}$ in frequency bins where foreground contamination is most significant. 

Moreover, we demonstrated the potential of BINGO to measure the dark energy equation of state. The joint \textit{Planck} 2018 + BINGO analysis yielded $w_0 = -0.71^{+0.22}_{-0.21}$ and $w_a = 0.0\pm 1.2$, consistent with the fiducial $\Lambda$CDM values ($w_0 = -1$, $w_a = 0$) at the $ 1.9\sigma$ confidence level. While \textit{Planck} provides high-precision measurements of most cosmological parameters, BINGO spans the low-redshift expansion history—the regime where dark energy dominates.  

In conclusion, this study quantifies the reliability and accuracy of our cosmological parameter constraints using BINGO simulations in conjunction with \textit{Planck} data.  By exploring uncertainties and testing multiple simulation realizations, we demonstrate BINGO’s promising potential for constraining cosmological parameters, HI history, and the dark energy equation of state.  However, our simulations simplify observational complexity: instrumental effects such as beam structure, 1/f noise, and polarization leakage require more realistic modeling in future work. This analysis represents a critical first step in validating BINGO’s capabilities. Next steps will prioritize improving simulations by addressing these systematics to strengthen BINGO’s role in probing cosmic acceleration.

Our analysis demonstrates improved cosmological constraints using BINGO simulations combined with the Planck 2018 likelihoods. However, two aspects of the fitting method could be refined in future work. First, the use of two HI-free parameters per frequency channel (totaling 60) introduces computational complexity and parameter degeneracies. Adopting a two-degree polynomial model for $\Omega_{\mathrm{HI}}$ and $b_{\mathrm{HI}}$ across BINGO’s redshift range \citep{zhang2022bingo} could reduce these to six parameters, significantly streamlining the analysis. Second, while the Gaussian approximation in our likelihood function performs adequately for the cases studied here, it may not fully capture non-Gaussian features inherent to the data. To address this, we aim to implement more robust likelihoods in future analyses.

\begin{acknowledgements}
The BINGO project is supported by FAPESQ-PB, the State of Paraiba, FINEP, and FAPESP-SP, Brazil, by means of several grants. The authors acknowledge the National Laboratory for Scientific Computing (LNCC/MCTI, Brazil) for providing HPC resources of the SDumont supercomputer, which have contributed to the research results reported within this paper. URL: http://sdumont.lncc.br. P.M.: This study was financed in part by the São Paulo Research Foundation (FAPESP) through grant 2021/08846-2 and 2023/07728-1. Also, this study was financed in part by the Coordenação de Aperfeiçoamento de Pessoal de Nível Superior – Brasil (CAPES) – Finance Code 001. F.B.A thanks the University of Science and Technology of China and the Chinese Academy of Science for grant
number KY2030000215. C.P.N. thanks Serrapilheira Institute and São Paulo Research Foundation
(FAPESP; grant 2019/06040-0) for financial support. E.A. is supported by a CNPq grant. E.J.M. thanks INPE for the financial support provided by the PCI fellowship. GAH acknowledges the funding from the Dean’s Doctoral Scholarship by the University of Manchester. L.O.P. acknowledges the grant 2023/07564-9, São Paulo Research Foundation (FAPESP). A.R.Q. work is supported by FAPESQ-PB. A.R.Q. acknowledges support by CNPq under process number 310533/2022-8. C.A.W. thanks CNPq for grants 407446/2021-4 and 312505/2022-1, the Brazilian Ministry of Science, Technology and Innovation (MCTI) and the Brazilian Space Agency (AEB) who supported the present work under the PO 20VB.0009.
\end{acknowledgements}

\bibliographystyle{aa}
\bibliography{bibliog}

\end{document}